\documentclass[a4paper,fleqn,usenatbib]{mnras}

\usepackage[T1]{fontenc}
\usepackage{ae,aecompl}
\usepackage{graphicx,longtable,multirow,amsmath,tabularx,lscape}
\usepackage{xspace}

\usepackage{natbib,bm}
\usepackage[usenames]{color}

\newcommand{\bd}{\begin{displaymath}}
\newcommand{\ed}{\end{displaymath}}
\newcommand{\be}{\begin{equation}}
\newcommand{\ee}{\end{equation}}
\newcommand{\beaa}{\begin{eqnarray*}}
\newcommand{\eeaa}{\end{eqnarray*}}
\newcommand{\bea}{\begin{eqnarray}}
\newcommand{\eea}{\end{eqnarray}}

\def\hequad{HE\,0435$-$1223}

\def\blens{B1608$+$656}
\def\rxjlens{RXJ1131$-$1231}

\def\Ok{\Omega_{\rm k}}

\def\Om{\Omega_{\rm m}}
\def\OL{\Omega_{\Lambda}}

\def\tdist{D_{\Delta t}}
\def\tdistmod{D_{\Delta t}^{\rm model}}
\def\Dd{D_{\rm d}}
\def\Dds{D_{\rm ds}}
\def\Ds{D_{\rm s}}

\def\kext{\kappa_{\rm ext}}
\def\gext{\gamma_{\rm ext}}

\def\zd{z_{\rm d}}
\def\zs{z_{\rm s}}

\def\hst{\textit{HST}}

\def\GLEE{{\sc Glee}\xspace}

\def\kms {\rm km\,s^{-1}}

\newcommand{\sref}[1]{Section~\ref{#1}}
\newcommand{\fref}[1]{Figure~\ref{#1}}

\newcommand{\eref}[1]{Equation~(\ref{#1})}

\newcommand{\zlens}{0.4546}
\newcommand{\zsrc}{1.689}
\newcommand{\dt}{2612_{-191}^{+208}~\mathrm{Mpc}}
\newcommand{\dtprec}{7.6}
\newcommand{\uh}{74.3_{-5.4}^{+6.0}~\mathrm{km~s^{-1}~Mpc^{-1}}}
\newcommand{\ulcdm}{73.1_{-6.0}^{+5.7}~\mathrm{km~s^{-1}~Mpc^{-1}}}
\newcommand{\combprec}{3.5}
\newcommand{\kextmed}{0.003}
\newcommand{\kextlo}{-0.016}
\newcommand{\kexthi}{0.034}
\newcommand{\rani}{r_{\rm ani}}
\newcommand{\reff}{r_{\rm eff}}
\newcommand{\appropto}{\mathrel{\vcenter{
  \offinterlineskip\halign{\hfil$##$\cr
    \propto\cr\noalign{\kern2pt}\sim\cr\noalign{\kern-2pt}}}}}

\title[HE0435 Lens Model]{H0LiCOW IV. Lens mass model of \hequad~and blind measurement of its time-delay distance for cosmology}

\author[K. C. Wong et al.]{\parbox{\textwidth}{
Kenneth C. Wong,$^{1,2,\dagger}$\thanks{E-mail: ken.wong@nao.ac.jp}
Sherry H. Suyu,$^{3,2,4}$
Matthew W. Auger,$^{5}$
Vivien Bonvin,$^{6}$
Frederic Courbin,$^{6}$
Christopher D. Fassnacht,$^{7}$
Aleksi Halkola,
Cristian E. Rusu,$^{7}$
Dominique Sluse,$^{8}$
Alessandro Sonnenfeld,$^{9}$
Tommaso Treu,$^{10}$
Thomas E. Collett,$^{11}$
Stefan Hilbert,$^{12,13}$
Leon V. E. Koopmans,$^{14}$
Philip J. Marshall,$^{15}$
and Nicholas Rumbaugh$^{7}$}
\\
\\
\parbox{\textwidth}{
$^{1}$National Astronomical Observatory of Japan, 2-21-1 Osawa, Mitaka, Tokyo 181-8588, Japan\\
$^{2}$Institute of Astronomy and Astrophysics, Academia Sinica (ASIAA), P.O.~Box 23-141, Taipei 10617, Taiwan\\
$^{3}$Max-Planck-Institut f{\"u}r Astrophysik, Karl-Schwarzschild-Str.~1, 85748 Garching, Germany\\
$^{4}$Physik-Department, Technische Universit\"at M\"unchen, James-Franck-Stra\ss{}e~1, 85748 Garching, Germany\\
$^{5}$Institute of Astronomy, University of Cambridge, Madingley Rd, Cambridge, CB3 0HA, UK\\
$^{6}$Laboratoire d'Astrophysique, Ecole Polytechnique F{\'e}d{\'e}rale de Lausanne (EPFL), Observatoire de Sauverny, CH-1290 Versoix, Switzerland\\
$^{7}$Department of Physics, University of California Davis, 1 Shields Avenue, Davis, CA 95616, USA\\
$^{8}$STAR Institute, Quartier Agora - All\'ee du six Ao\^ut, 19c B-4000 Li\`ege, Belgium\\
$^{9}$Kavli Institute for the Physics and Mathematics of the Universe (Kavli IPMU, WPI), University of Tokyo, Chiba 277-8583, Japan\\
$^{10}$Department of Physics and Astronomy, University of California, Los Angeles, CA 90095-1547, USA\\
$^{11}$Institute of Cosmology and Gravitation, University of Portsmouth, Burnaby Rd, Portsmouth PO1 3FX, UK\\
$^{12}$Exzellenzcluster Universe, Boltzmannstr. 2, 85748 Garching, Germany\\
$^{13}$Ludwig-Maximilians-Universit{\"a}t, Universit{\"a}ts-Sternwarte, Scheinerstr. 1, 81679 M{\"u}nchen, Germany\\
$^{14}$Kapteyn Astronomical Institute, University of Groningen, PO Box 800, NL-9700 AV Groningen, The Netherlands\\
$^{15}$Kavli Institute for Particle Astrophysics and Cosmology, Stanford University, 452 Lomita Mall, Stanford, CA 94035, USA\\
$^{\dagger}$EACOA Fellow
}}

\date{Accepted XXX. Received YYY; in original form ZZZ}
\pubyear{2016}

\begin{document}
\label{firstpage}
\pagerange{\pageref{firstpage}--\pageref{lastpage}}
\maketitle

\begin{abstract}
Strong gravitational lenses with measured time delays between the
multiple images allow a direct measurement of the time-delay
distance to the lens, and thus a measure of cosmological
parameters, particularly the Hubble constant, $H_{0}$.  We present a blind lens model analysis of the quadruply-imaged quasar
lens \hequad~using deep {\it Hubble Space Telescope} imaging,
updated time-delay measurements from the COSmological MOnitoring of
GRAvItational Lenses (COSMOGRAIL), a measurement of the velocity
dispersion of the lens galaxy based on Keck data, and a characterization of the mass
distribution along the line of sight.  \hequad~is the third lens analyzed as a part of the $H_{0}$ Lenses in COSMOGRAIL's Wellspring (H0LiCOW) project.  We account for various
sources of systematic uncertainty, including the detailed treatment
of nearby perturbers, the parameterization of the galaxy light and
mass profile, and the regions used for lens modeling.  We constrain
the effective time-delay distance to be $\tdist = \dt$, a precision of \dtprec\%.  From \hequad~alone, we infer a Hubble constant of $H_{0} = \ulcdm$
assuming a flat $\Lambda$CDM cosmology.  The cosmographic inference based on the three lenses analyzed by H0LiCOW to date is presented in a companion paper (H0LiCOW Paper V).
\end{abstract}


%-------------------------------------------------------------------------------

\section{Introduction} \label{sec:intro}

The flat $\Lambda$CDM cosmological model is the concordance
model of our Universe today.  It is consistent with a variety of
independent experiments, including an analysis of the cosmic microwave
background (CMB) by the {\it Planck} mission \citep{planck2015}.  The
{\it Planck} results provide the most precise cosmological parameter
constraints to date, under the assumption of spatial flatness.  However, there is no physical reason to assume flatness, and if the
flatness assumption is relaxed, there are strong degeneracies among
the cosmological parameters inferred from CMB data, particularly with the Hubble constant,
$H_{0}$ \citep[e.g.,][]{freedman2012,riess2016}.  Therefore, an independent determination of $H_{0}$ is
crucial for understanding the nature of the Universe \citep[e.g.,][]{hu2005,suyu2012c,weinberg2013}.

The idea of using gravitational lens time delays to measure the Hubble
constant dates back to \citet{refsdal1964}.  In practice, gravitational
lens time delays provide a one-step method to determine the distance and hence the Hubble
constant
\citep[e.g.,][]{vanderriest1989,keeton1997,schechter1997,kochanek2003,koopmans2003,saha2006,oguri2007,fadely2010,suyu2010b,suyu2013,sereno2014,rathnakumar2015,birrer2016,chen2016}.  This method is independent of the cosmic distance ladder \citep[e.g.,][]{riess2011,freedman2012} and serves as a key test of possible systematic effects in individual $H_{0}$ probes.  This method rests on the
fact that light rays emitted from the source at the same instant will
take different paths through spacetime at each of the image positions.
These paths have different lengths and traverse different
gravitational potentials before reaching the observer, leading to an
offset in arrival times.  If the source exhibits variations in its
flux, the delays can be measured by monitoring the lensed images.  The
measured time delays can be used to calculate the time-delay
distance, a combination of angular diameter distances among the
observer, lens, and source.  The time-delay distance is primarily sensitive to $H_{0}$, with weaker dependence on other cosmological
parameters \citep[e.g.,][]{coe2009,treu2016}.

However, a precise and accurate determination of $H_{0}$ through this
method requires a variety of observational data.  A dedicated
long-term monitoring campaign is necessary to obtain accurate time
delays, as the uncertainty in $H_{0}$ is directly related to the
relative uncertainty in the measured time delays.  Deep,
high-resolution imaging is required to accurately model the lens using
the extended source images, which is needed to break degeneracies
between the mass profile and the underlying cosmology
\citep[e.g.,][]{kochanek2002,warren2003}.  In order to reduce the effects of the mass sheet degeneracy \citep[e.g.,][]{falco1985,gorenstein1988,saha2000,schneider2013,xu2016}, a measurement of the lens galaxy's velocity dispersion \citep[e.g.,][]{koopmans2003,koopmans2004} and an estimate of the external convergence, $\kext$, along the line of sight (LOS) is needed.
$\kext$ can also bias the lens
model parameters if unaccounted for \citep[e.g.,][]{collett2013,greene2013,mccully2014,mccully2016}.

In an effort to provide an accurate independent estimate of $H_{0}$
using time-delay lenses, we use a number of new datasets as part of our
project, $H_{0}$ Lenses in COSMOGRAIL's Wellspring (H0LiCOW), to model five lensed quasars.  These datasets
include high-resolution imaging with the {\it Hubble Space Telescope}
(\hst), precise time-delay measurements from the COSmological
MOnitoring of GRAvItational Lenses \citep[COSMOGRAIL;][]{courbin2005,eigenbrod2005,bonvin2016b}
project and from Very Large Array (VLA) monitoring \citep{fassnacht2002}, a photometric and spectroscopic survey to characterize
the LOS mass distribution to estimate $\kext$ in these systems, and stellar velocity dispersion measurements of the strong lens galaxies.  With five separate lenses, we plan to account for systematic
uncertainties and obtain a robust constraint on $H_{0}$ to $< \combprec\%$
precision.

In this paper, we present the results of a detailed lens modeling
analysis of the gravitational lens \hequad~using new high-resolution
imaging data from \hst.  \hequad~is the third H0LiCOW system analyzed
in this manner, following \blens~\citep{suyu2010b} and \rxjlens~\citep{suyu2013, suyu2014}.  This paper is the fourth in a series of papers detailing our analysis of \hequad.  The other papers include an overview of the H0LiCOW project \citep[][hereafter H0LiCOW Paper I]{suyu2016}, a spectroscopic survey of the \hequad~field and a characterization of the groups along the LOS \citep[][hereafter H0LiCOW Paper II]{sluse2016}, a photometric survey of the \hequad~field and an estimate of $\kext$ due to the external LOS structure \citep[][hereafter H0LiCOW Paper III]{rusu2016}, and a presentation of our latest time-delay measurements for \hequad~and the cosmological inference from our combined analysis of \hequad, \blens, and \rxjlens~\citep[][hereafter H0LiCOW Paper V]{bonvin2016a}.

This paper is organized as follows.  We provide a brief overview of
using time-delay lenses for cosmography in \sref{sec:theory}.  In
\sref{sec:data}, we describe the observational data used in our
analysis.  We describe our lens modeling procedure in
\sref{sec:lensmod}.  The time-delay distance results and their implications for cosmology are presented in
\sref{sec:results}.  We summarize our main conclusions in
\sref{sec:conclusions}.  Throughout this paper, all magnitudes given
are on the AB system.

%-------------------------------------------------------------------------------

\section{Time-Delay Cosmography} \label{sec:theory}

\subsection{Time-delay distance} \label{subsec:tddist}
When a source is gravitationally lensed, the light travel time from
the source to the observer depends on both the path length of the
light rays and the gravitational potential of the lens through which
the rays pass.  For a single lens plane, the excess time delay of an image at an
angular position $\bm{\theta} = (\theta_{1}, \theta_{2})$ with
corresponding source position $\bm{\beta} = (\beta_{1}, \beta_{2})$
relative to the case of no lensing is
\begin{equation} \label{eq:td}
t(\bm{\theta}, \bm{\beta}) = \frac{\tdist}{c} \left[ \frac{(\bm{\theta} - \bm{\beta})^{2}}{2} - \psi(\bm{\theta}) \right],
\end{equation}
where $\tdist$ is the time-delay distance and
$\psi(\bm{\theta})$ is the lens potential.  The time-delay distance \citep{refsdal1964,schneider1992,suyu2010b} is
defined\footnote{For historical reasons, the time-delay distance is written in terms of angular diameter distances.  A more natural definition is $\tdist \equiv {\hat{D}_{\mathrm{d}} \hat{D}_{\mathrm{s}}}/{\hat{D}_{\mathrm{ds}}}$ where $\hat{D}$  are the proper distances that the photons have travelled.} as
\begin{equation} \label{eq:ddt}
\tdist \equiv (1+\zd) \frac{\Dd \Ds}{\Dds},
\end{equation}
where $\zd$ is the lens redshift, $\Dd$ is the angular diameter
distance to the lens, $\Ds$ is the angular diameter distance to the
source, and $\Dds$ is the angular diameter distance between the lens
and the source.  Since $\tdist$ has units of distance, it is inversely
proportional to $H_{0}$.

For lens systems with multiple deflectors at distinct redshifts, the
observed time delays depend on various combinations of the angular
diameter distances measured between us, the multiple deflectors, and
the source, and the observed time delays are no longer proportional to a single time-delay distance.  The observed image positions depend on the multi-plane
lens equation
\citep[e.g.,][]{blandford1986,kovner1987,schneider1992,petters2001,collett2014,mccully2014}.
However, for a system where the lensing is dominated by a single
plane, the observed time delays are primarily sensitive to the
time-delay distance defined in \eref{eq:ddt}, with the
deflector redshift as that of the primary strong lens plane.  We show
in Section~\ref{subsec:cosmo} that this approximation is valid for
\hequad~and our results can thus be interpreted as a constraint on
$\tdist(\zd,\zs)$, which we refer to as the effective time-delay
distance measured by this system.  Hereafter, $\tdist$ refers to this
effective time-delay distance unless otherwise indicated.

For variable sources such as active galactic nuclei (AGN), it is possible to monitor the fluxes
of the lensed images at positions $\bm{\theta}_{i}$ and
$\bm{\theta}_{j}$ and measure the time delay, $\Delta t_{ij} \equiv
t(\bm{\theta}_{i}, \bm{\beta}) - t(\bm{\theta}_{j}, \bm{\beta})$,
between them \citep[e.g.,][]{vanderriest1989,schechter1997,fassnacht1999,fassnacht2002,kochanek2006,courbin2011}.  The lens potentials at the two image positions,
$\psi(\bm{\theta}_{i})$ and $\psi(\bm{\theta}_{j})$, as well as the
source position, $\bm{\beta}$, can be determined from a mass model of
the system.  Therefore, lenses with measured time delays and accurate
lens models can be used to constrain $\tdist$.

A complicating factor in using time-delay lenses for cosmography is
the fact that all mass along the LOS contributes to the lens
potential that the light rays pass through.  These external perturbers
not only affect the lens model of the system, but also lead to
additional focusing and defocusing of the light rays, which in turn
affects the measured time delays \citep[e.g.,][]{seljak1994}.  If unaccounted for, these external
perturbers can lead to biased inferences of $\tdist$.  If
effects of LOS perturbers are small, they can be
approximated by an external convergence term in the lens plane,
$\kext$ \citep[neglecting the $1-\beta$ terms that enter into a more accurate prescription;][]{keeton2003,mccully2014}.  The true $\tdist$ is related to the
$\tdistmod$ inferred from a mass model that does not account for $\kext$ by
\begin{equation} \label{eq:ddtkappa}
\tdist = \frac{\tdistmod}{1-\kext}.
\end{equation}
$\kext$ cannot be constrained from the lens model due
to the mass sheet degeneracy
\citep[e.g.,][]{falco1985,gorenstein1988,saha2000}, in which the
addition of a uniform mass sheet and a rescaling of the source plane
coordinates can affect the product of the time delays and $H_{0}$ but
leave other observables unchanged.

The above degeneracy caused by $\kext$ can be broken or substantially mitigated by estimating the mass distribution along the LOS \citep[e.g.,][]{fassnacht2006,momcheva2006,momcheva2015,williams2006,wong2011}. However, for perturbers that are very massive or projected very close to the lens, they may need to be included explicitly in the mass model, as their higher-order effects need to be properly accounted for \citep{mccully2016}.  On the other hand, the lens profile is also degenerate with the time-delay distance in that the radial profile slope is tightly correlated with the time-delay distance \citep[e.g.,][]{kochanek2002,wucknitz2002,suyu2012a}.  The profile degeneracy affects models that share the same form of mass density profile (e.g., a power-law density profile), as well as models with different density profiles (described analytically or not).  Furthermore, the profile degeneracy can mimic the effects of the mass sheet degeneracy since different profiles can exactly or approximately be mass sheet transformations of one form or another \citep[e.g.,][]{schneider2013,schneider2014,unruh2016}.  With reasonable assumptions about the lens galaxy's mass profile, these degeneracies can be reduced by augmenting the lensing data with stellar kinematics measurements of the lens galaxy \citep[e.g.,][]{treu2002,koopmans2003,auger2010,suyu2014}.  Including the velocity dispersion in the modeling helps to constrain any internal uniform mass component from a local galaxy group that the dynamics is sensitive to \citep{koopmans2004}.

\subsection{Joint Inference} \label{subsec:joint}
Our inference of $\tdist$ follows that of \citet{suyu2013}, but with some
important modifications.  Our
observational data sets are denoted by $\bm{d_{\mathrm{HST}}}$ for the
\hst~imaging data, $\bm{\Delta t}$ for the time delays, $\sigma$ for the
velocity dispersion of the lens galaxy, and $\bm{d_{\mathrm{LOS}}}$ for
the properties of the LOS mass distribution determined from our
photometric and spectroscopic data. We want to obtain the posterior
probability distribution function (PDF) of the model parameters
$\bm{\nu}$ given the data, $P(\bm{\xi} | \bm{d_{\rm HST}, \Delta t, \sigma,
d_{\mathrm{LOS}},A})$.  The vector~$\bm{\xi}$ includes the lens model
parameters~$\bm{\nu}$,
the cosmological parameters $\bm{\pi}$ (Section~\ref{subsec:cosmo}),
and nuisance parameters representing the external convergence ($\kext$; Section~\ref{subsec:convergence}) and anisotropy radius ($r_{\rm ani}$; Section~\ref{subsec:kinematics}),
each of which we introduce and discuss in the sections indicated.
${\bm A}$ denotes a discrete set of assumptions we make about the form of the
model, including the choices we have to make about the data modeling region, the set-up of the source reconstruction grid, the treatment of
the various deflector mass distributions, etc.  In general, ${\bm A}$ cannot be fully captured by continuous parameters.
By Bayes' theorem, we have that
\begin{eqnarray} \label{eq:pdf}
&& P(\bm{\xi} | \bm{d_{\rm HST}, \Delta t, \sigma, d_{\mathrm{LOS}}, A}) \nonumber \\
        &\propto& P(\bm{d_{\rm HST}, \Delta t, \sigma, d_{\mathrm{LOS}}} | \bm{\xi,A}) P(\bm{\xi} | \bm{A}),
\end{eqnarray}
where ${P(\bm{d_{\rm HST}, \Delta t, \sigma, d_{\mathrm{LOS}}}} | \bm{\xi,
A})$ is the joint likelihood function and $P(\bm{\xi} | \bm{A})$ is the
prior PDF for the parameters given our assumptions.  Since the data sets
are independent, the likelihood can be separated,
\begin{eqnarray} \label{eq:pdf_sep}
P(\bm{d_{\rm HST}, \Delta t, \sigma, d_{\mathrm{LOS}}} | \bm{\xi, A}) &=& P(\bm{d_{\rm HST}} | \bm{\xi, A}) \nonumber \\
  &&  \times P(\bm{\Delta t} | \bm{\xi, A}) \nonumber \\
  &&  \times P(\bm{\sigma} | \bm{\xi, A})  \nonumber \\
  &&  \times P(\bm{d_{\mathrm{LOS}}} | \bm{\xi, A}).
\end{eqnarray}
We note that \eref{eq:pdf_sep} assumes the approximation that
the LOS can be decoupled from the lens model.  We can calculate the
individual likelihoods separately and combine them as in
\eref{eq:pdf_sep} to get the final posterior PDF for a given
set of assumptions.

In Section~\ref{subsec:sys_tests}, we lay out a range of systematics
tests where we vary the content of $\bm{A}$ and repeat the inference
of ${\bm \xi}$. Such a sensitivity analysis is important for checking
the magnitude of various known but unmodeled systematic effects, but it
leaves us with the question of how to combine the results. We note that
the marginalization integral over these assumptions can be approximated
as a sum as follows (denoting all four datasets by $\bm{d}$),
\begin{eqnarray}
    P(\bm{\xi} | \bm{d}) &=& \int P(\bm{\xi} | \bm{d, A}) P(\bm{A} | \bm{d})\, d\bm{A}  \nonumber \\
                       &\appropto& \sum_k P(\bm{\xi} | \bm{d, A_k}) P(\bm{d} | \bm{A_k}) \nonumber \\
                       &\appropto& \sum_k P(\bm{\xi} | \bm{d, A_k}), \label{eq:combination}
\end{eqnarray}
provided the following two statements are true: first, that the
prior PDF over possible assumptions is uniform, and that our sampling
of possible assumptions is fair. We choose reasonable
variations in the systematic effects to try to achieve this. The second
is that the evidence $P(\bm{d} | \bm{A_k})$ does not change appreciably between
inferences; this is likely to be true if the goodness of fit does not change,
and the parameter priors and volumes are not very different.  Under these
assumptions, \eref{eq:combination} shows that a {\it sum}
of the posterior PDFs is an approximation to the posterior PDF marginalized
over the tested systematic effects.

%-------------------------------------------------------------------------------

\section{Data} \label{sec:data}
\hequad~(J2000:
$4^{\mathrm{h}}38^{\mathrm{m}}14\mbox{\ensuremath{.\!\!^{\mathrm{s}}}}9$,
$-12^{\circ}17\arcmin14\farcs4$) is a quadruply-lensed quasar
discovered by \citet{wisotzki2002} as part of the Hamburg/ESO survey
for bright QSOs \citep{wisotzki2000}.  The main deflector is a massive
elliptical galaxy at a redshift of $\zd = \zlens \pm 0.0002$
\citep{morgan2005}, and the source redshift is $\zs = \zsrc$\footnote{We note that \citet{sluse2012} measure an updated source redshift of $\zs = 1.693$.  We use the original value of $\zs = \zsrc$ in our analysis but verify that using this updated measurement does not impact our results.}.  Our
spectroscopic observations reveal that the lens is part of a galaxy
group with a velocity dispersion of $\sigma = 471 \pm 100~\mathrm{km~s}^{-1}$ measured from 12 member galaxies (H0LiCOW Paper II), which is independently confirmed
by \citet{wong2011} and Wilson et al. (in preparation) based on a spectroscopic study by \citet{momcheva2006,momcheva2015}.  We
present the \hst~imaging used for lens modeling in \sref{subsec:hst},
the time delays measured by COSMOGRAIL in \sref{subsec:td}, the
spectroscopy of the lens galaxy for measuring the lens stellar
velocity dispersion in \sref{subsec:vddata}, and ground-based imaging
and spectroscopy to characterize the lens environment in
\sref{subsec:lensenv}.

\subsection{HST Imaging} \label{subsec:hst}
We obtain deep \hst~observations of \hequad~using the Wide Field
Camera 3 (WFC3) IR channel in the F160W band (Program \#12889; PI:
Suyu).  The details of these observations are presented in H0LiCOW Paper I, which we summarize here.  
Using a combination of short (44 s) and long (599 s) exposures, we 
reconstruct the brightness distribution of both the lensed AGN
and host galaxy.
  We reduce the images using {\sc
  DrizzlePac}\footnote{{\sc DrizzlePac} is a product of the Space
  Telescope Science Institute, which is operated by AURA for
  NASA.}. The images are drizzled to a final pixel scale of
$0\farcs08$ without masking the bright AGN pixels, as they are well
characterized.

We also use archival observations from the Advanced Camera for Surveys
(ACS) on \hst~in the F555W and F814W filters (Program \#9744; PI:
Kochanek).  
  The images are reduced using {\sc
  MultiDrizzle}\footnote{MultiDrizzle is a product of the Space
  Telescope Science Institute, which is operated by AURA for NASA.}
with charge transfer inefficiency taken into account \citep[e.g.,][]{anderson2010,massey2010}.  The final pixel scale of the reduced
images is 0\farcs05.

\begin{figure*}
\includegraphics[width=\textwidth]{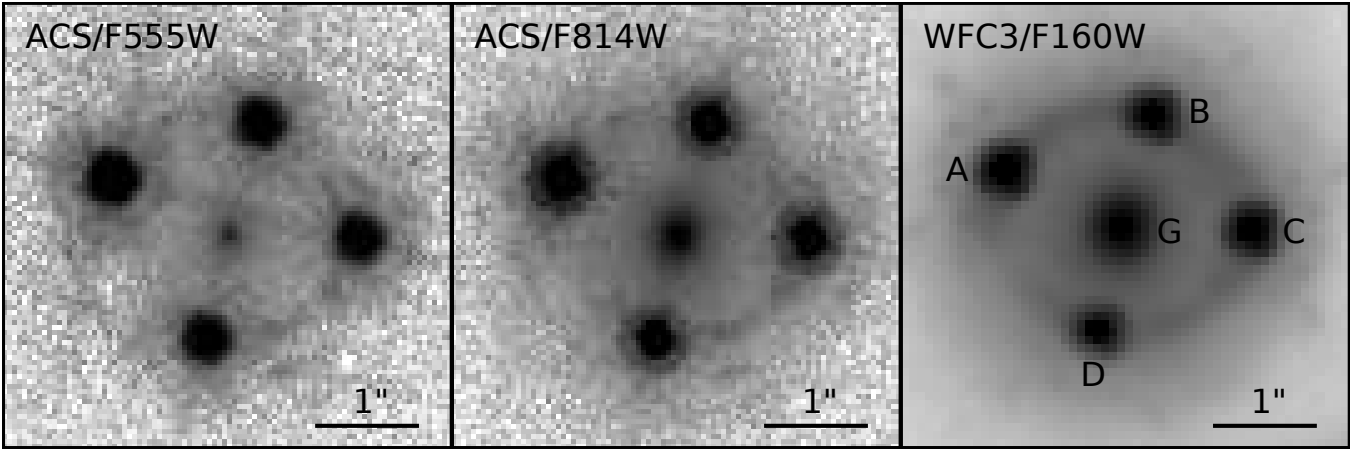}
\caption{ \hst~images of \hequad.  Shown are cutouts of the lens
  system used for lens modeling in the ACS/F555W (left), ACS/F814W
  (middle), and WFC3/F160W (right) bands.  The images are $4\farcs5$
  on a side.  The scale is indicated in the bottom right of each
  panel.  The main lens galaxy (G) and lensed quasar images (A, B, C,
  D) are marked.}
\label{fig:images}
\end{figure*}

We create cutouts of the \hst~images around the lens and define an arcmask in each band in which we perform the modeling.  For the ACS bands, we use a $90 \times 90$
pixel cutout ($4\farcs5$ on a side), and for the WFC3/F160W band, we
use a $60 \times 60$ pixel cutout ($4\farcs8$ on a side).  These
cutouts are shown in Figure~\ref{fig:images}.

To generate the initial point spread function (PSF) of the exposures,
we first select three stars in the field which are close to the
lens galaxy in angular separation to minimize CCD distortion effects,
and which have approximately the same brightness as the lensed AGN
images to avoid any PSF broadening effects.  We then simultaneously
fit these stars with a Moffat profile plus a regularized fine-pixel
array.  The exposures are sky-subtracted prior to the PSF fitting.
The details of this fitting procedure are described by \citet{cantale2016a} and are based on ideas presented in
\citet{magain1998}.  A successful application of the procedure is presented by
\citet{cantale2016b}.  We then use this initial PSF as the starting
point for our iterative PSF correction procedure (see
Section~\ref{subsec:overview}).

The weight images are constructed as follows.  We take a large,
relatively sparse area of the image and approximate the background
noise as the normalized median absolute deviation (NMAD), defined as
$\mathrm{NMAD} \equiv 1.48 \times \mathrm{median}(|p_{i} -
\mathrm{median}(p)|)$, where $p_{i}$ is the value of pixel $i$ and
median($p$) is the median of all pixels in the selected area.  We use the NMAD, which is a good approximation to the standard deviation, as it is less sensitive to outliers.  We create a
``noise image" that has the same dimensions as the lens galaxy cutout
with all pixels initialized to the value of the background noise.  We
then add Poisson noise to this noise image by taking all pixels in the
lens galaxy cutout where the flux is greater than the background noise level and adding in quadrature the
square root of each pixel value (normalized by its effective exposure
time) to the corresponding pixel in the noise image (this is because the units of the science image are counts per second).  The noise image is then squared and inverted to obtain the weight image.  We note that while the background noise for the WFC3 IR camera depends on the number of non-destructive reads, we verify that the number of reads in the region of the lensed arc is the same as for the blank sky patch used for estimating the background noise, so this procedure is valid.

When modeling with these weights, there are large residuals near the AGN image centers due to our inability to model the PSF on a grid of pixels with sufficient accuracy.  This can lead to biased results as the model will be influenced by these relatively small areas rather than the large-scale features of the source, so we compensate for this by reducing the weight in these regions \citep[e.g.,][]{suyu2012a}.  We scale the weight in these regions by a power law such that a pixel originally given a noise
value of $p_{i}$ is rescaled to a noise value of $A \times p_{i}^{b}$.  The $A$ and $b$ are constants that are different for each band and are chosen such that the normalized residuals in the AGN image regions are
approximately consistent with the normalized residuals in the rest of
the arc region.

We note that in determining the effective exposure time on a
pixel-by-pixel basis, we turn off the bad pixel masking in a
$3\times3$ pixel region around the brightest pixel of each of the
lensed AGN images.
This is done because allowing bad pixel masking results in
interpolations of the image pixels that cause the four AGN images to
exhibit different PSF profiles, which complicates our iterative PSF
correction scheme (Section~\ref{subsec:overview}).
Turning off the bad pixel mask produces more faithfully and
consistently the four AGN images.  Since the majority of the lens mass
model constraints comes from the lensing arcs away from the centers of
the AGN images, we have checked that these arcs do not have bad pixels
that would affect our lens mass model.

\subsection{Time-delay measurements} \label{subsec:td}
Time-delay measurements for \hequad~were initially given in \citet{courbin2011}.  Further monitoring of the system by COSMOGRAIL has since improved the time delay accuracy and precision, completing the data from \citet{courbin2011} with $\sim$1300 exposures of 6 min each for a total of 301 new observing nights ranging from 2010 September to 2016 April. The details of the data acquisition and time-delay measurements used in
our analysis are presented in H0LiCOW Paper V, but we
summarize the main results here.

The data treatment follows the procedure described by
\citet{tewes2013b}. Each observing epoch is corrected following the
standard reduction steps (bias subtraction, flat-fielding and sky
correction). The PSF is estimated following the procedure described in
Section~\ref{subsec:hst}. The exposures are then normalized using bright,
non-saturated stars in the field of view. The photometry of the four
images of \hequad~is obtained on each exposure using the \citet{magain1998} deconvolution photometry presented in \citet{cantale2016b}. The light
curves obtained with this method are presented in Figure 2 of H0LiCOW Paper V.

The measurement of the time delays between each pair of images follows
the formalism introduced by \citet{tewes2013a}. The common intrinsic
variability of the quasar and the four independent extrinsic variability curves
are fitted using free-knot splines. The curves are then shifted in
time to optimize the fit. The uncertainties on the time-delay
measurements are estimated using a Monte Carlo approach.
A set of 1000 synthetic light curves are drawn, mimicking the light curves and the time delay constraining power of the observed data \citep{tewes2013a}. It is important that the synthetic datasets span a range of plausible true time delays, as this allows us to verify that the estimator accurately responds to theses input delays \citep[i.e., does not suffer from lethargy, described in][]{rathnakumar2015} and has not been involuntarily fine-tuned to recover a particular value of the time delay.
Various tests on the data reduction process
and curve-shifting technique have been performed successfully to
ensure the reliability of the time-delay measurements. We use the time
delays relative to image A: $\Delta t_{\mathrm{AB}} = -8.8 \pm 0.8$
d, $\Delta t_{\mathrm{AC}} = -1.1 \pm 0.7$ d, and $\Delta
t_{\mathrm{AD}} = -13.8 \pm 0.9$ d, where the uncertainties represent 1$\sigma$ confidence intervals.

\subsection{Stellar velocity dispersion of lens galaxy}
\label{subsec:vddata}
\begin{figure*}
\includegraphics[width=\textwidth]{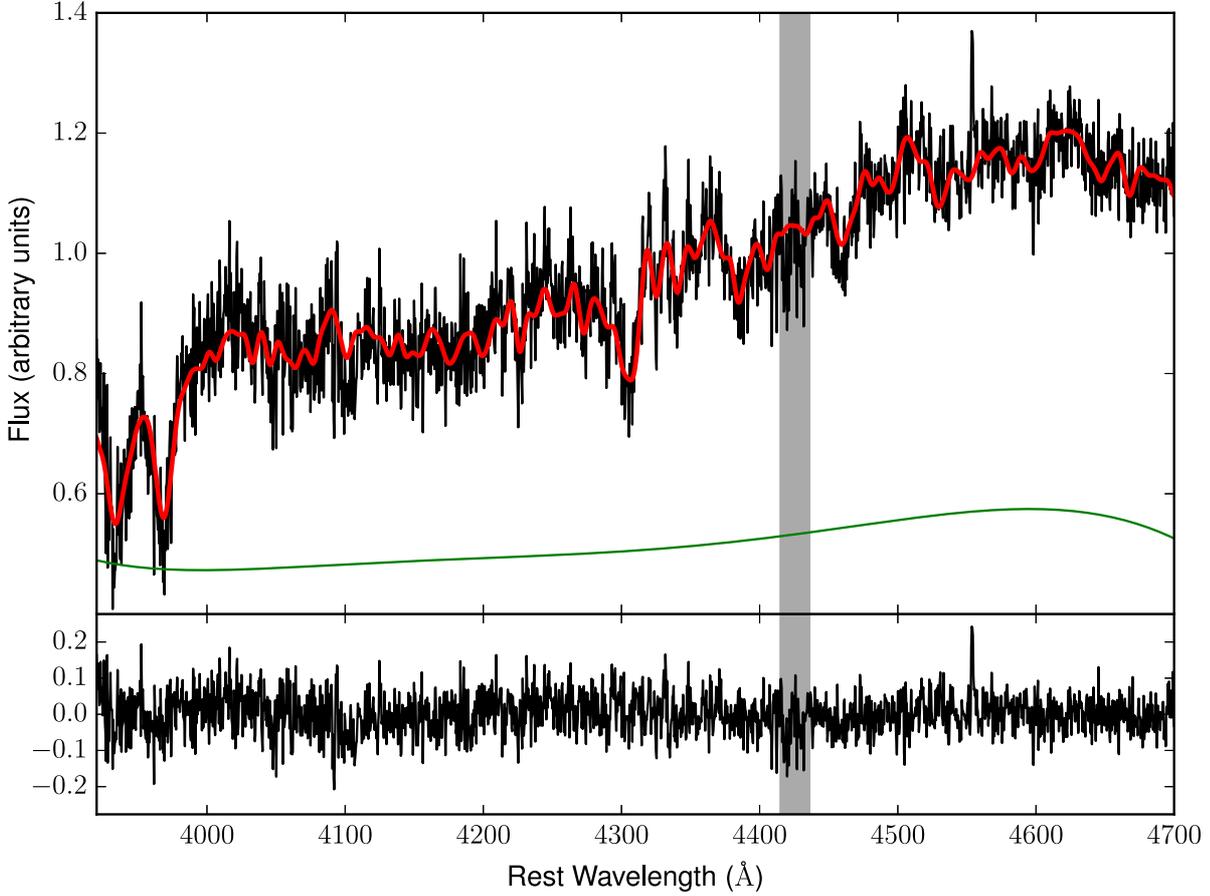}
\caption{
{\bf Top:} Keck/LRIS spectrum of \hequad~with the best-fitting model overplotted in red and a polynomial continuum, which accounts for contamination from the lensed QSO images and template mismatch, shown in green.  We find that $\sigma = 222\pm15$~km~s$^{-1}$, including systematic uncertainties due to the templates used, the region of the spectrum that was fitted, and the order of the polynomial continuum. The grey vertical band represents a wavelength range that is excluded from the fit due to the presence of a strong Mg II absorption system. {\bf Bottom:} Residuals from the best fit.}
\label{fig:spectrum}
\end{figure*}

\hequad~was observed with the Low Resolution Imaging Spectrometer \citep[LRIS;][]{oke1995} on the Keck I telescope on 2011 January 4. Six exposures of 1200 s were obtained in 0\farcs8 seeing with the red arm of the spectrograph using the 831/8200 grating, which has a dispersion of 0.58~\AA~pixel$^{-1}$ and yields an effective resolution $\sigma_{\mathrm{res}} \sim 37$~km~s$^{-1}$. The 0\farcs75 slit was oriented to intersect the eastern- and western-most lensed QSO images (i.e., at a position angle of $76^{\circ}$) and a 4-pixel (0\farcs54) aperture was used to extract one-dimensional spectra from each exposure. These six spectra were then resampled to a single spectrum using spline interpolation and rejecting pixels affected by cosmic rays or other artifacts; the resulting spectrum is shown in Figure~\ref{fig:spectrum}. The velocity dispersion was obtained following the same procedure as in \citet{suyu2010b,suyu2013}, resulting in an inference of $\sigma = 222$~km~s$^{-1}$ with a statistical uncertainty of $11~\mathrm{km~s^{-1}}$ and a systematic uncertainty of $\sim 10$~km~s$^{-1}$ due to the templates used, the region of the spectrum that was fitted, and the order of the polynomial continuum.  We therefore adopt an overall uncertainty of $\sigma_\sigma = 15$~km~s$^{-1}$.  This measurement is in agreement with a previous determination of $\sigma = 222 \pm 34$~km~s$^{-1}$ by \citet{courbin2011} within a 1\arcsec~aperture.

\subsection{Lens Environment: Photometry and Spectroscopy} \label{subsec:lensenv}
To account for the effects of LOS structure, we have obtained deep
multi-band photometry and multi-object spectroscopy in the
\hequad~field to characterize the external mass distribution.  Details
of the photometric observations and inference on
$\kext$ are presented in H0LiCOW Paper III, and the details of the spectroscopic data are presented in H0LiCOW Paper II, but we summarize the data here.

Our wide-field photometric data consist of 
ground-based $ugriJHKs$ observations, as well as 3.6, 4.5, 5.8, and 8.0 $\mu$m observations with the {\it Spitzer} Infrared Array Camera \citep[IRAC;][]{fazio2004}.  
We infer photometric redshifts and stellar masses using PSF-matched photometry measured with {\sc SExtractor}.  We use {\sc Lephare} \citep{ilbert2006} to measure stellar masses for the best-fitting redshift using the spectral energy distribution (SED) templates employed by CFHTLenS \citep{velander2014}, which assume a \citet{chabrier2003} initial mass function.

\begin{figure}
\includegraphics[width=0.5\textwidth]{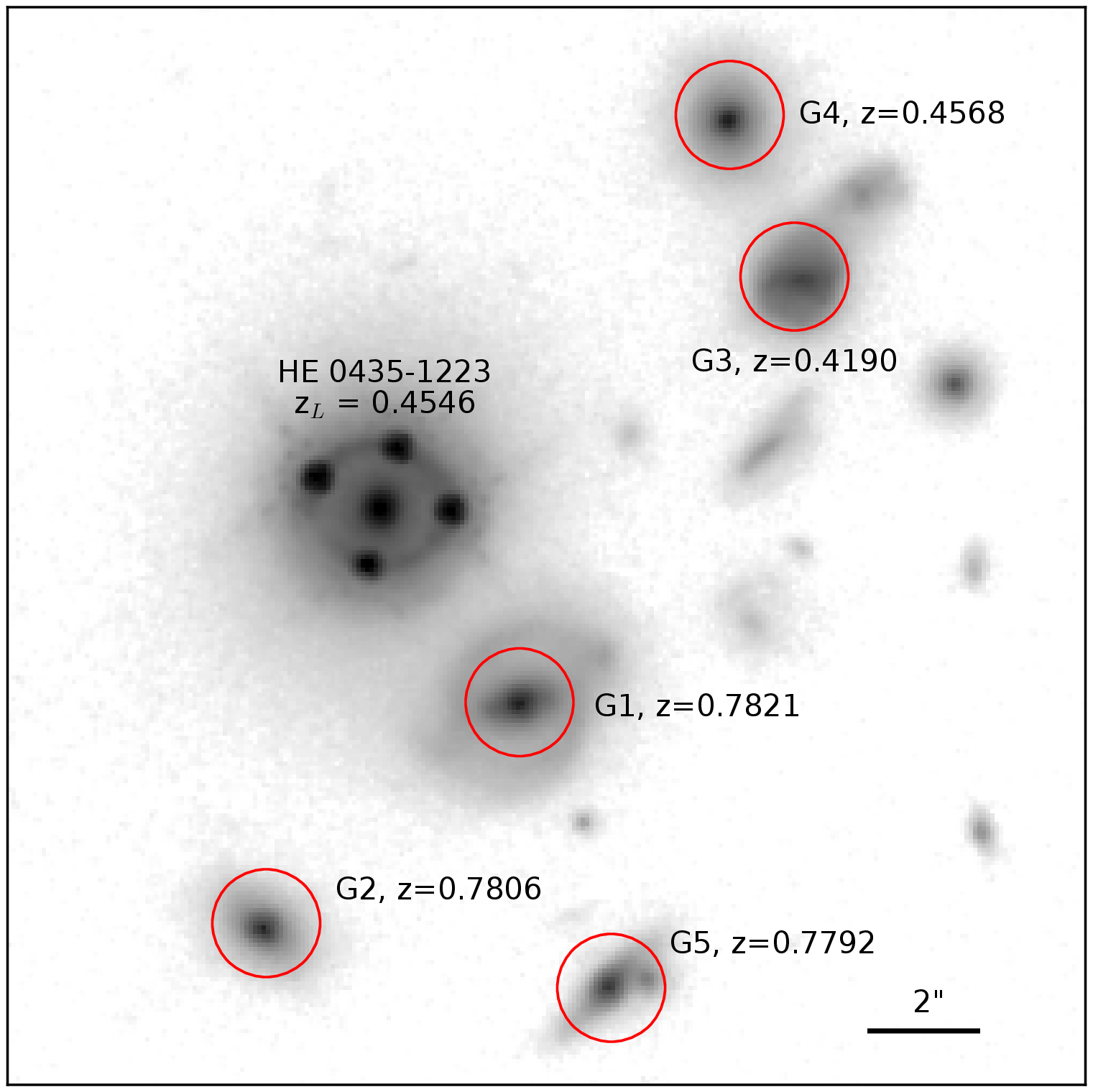}
\caption{ \hst/WFC3 F160W image of a $20\arcsec \times 20\arcsec$
  field around \hequad.  The angular scale is indicated in the bottom
  right corner.  The five most significant nearby perturbers are marked with red circles,
  and the redshifts of the perturbers are indicated.  G1 is included
  explicitly in our model, as it is the most massive and nearest in
  projection to \hequad.  We also test the effects of including the
  other perturbers as one of our systematics tests.
\label{fig:field}}
\end{figure}

The wide-field spectroscopic data are taken with a combination of Keck/LRIS, the Focal Reducer/low-dispersion Spectrograph 2 \citep[FORS2;][]{appenzeller1998} on the Very Large Telescope (VLT), and the Gemini Multi-Object Spectrograph \citep[GMOS;][]{hook2004}, and 
are combined with existing spectroscopic observations of this field \citep{momcheva2006,momcheva2015}.
It is particularly important to model the most significant perturbers, as their effects
may not be adequately accounted for by external shear alone
\citep{mccully2014}.  \citet{mccully2016} find that the most
significant perturbers are those that are massive, projected close to
the lens, and that are in the foreground of the lens redshift.  H0LiCOW Paper II presents an estimate of the relative significance of nearby perturbers to \hequad~as quantified by their flexion shift, $\Delta_{3} x$ \citep{mccully2016}, and finds that at most, the five nearest perturbers should be accounted for explicitly, with all other perturbers having a negligible influence.  Figure~\ref{fig:field} shows the lens and the relative positions and redshifts of these five perturbers, all brighter than $i = 22.5$ mag and projected within 12\arcsec~of the lens.

%-------------------------------------------------------------------------------

\section{Lens Modeling} \label{sec:lensmod}
In this section, we describe our procedure to simultaneously model the
images in the three \hst~bands and the time delays to infer the lens
model parameters.

\subsection{Overview} \label{subsec:overview}
We perform our lens modeling using \GLEE, a software package developed by
S.~H.~Suyu and A.~Halkola \citep{suyu2010a,suyu2012b}.  The lensing
mass distribution is described by a parameterized profile.  The
extended host galaxy of the source
is modeled separately on a $40\times40$ pixel grid with curvature regularization \citep{suyu2006}.
The lensed
quasar images are modeled as point sources convolved with the PSF.
By modeling the quasar images on the image plane
independently from the extended host galaxy light distribution, we
allow for variations in quasar fluxes due to microlensing, time delays, and
substructure.  The lens galaxy light distribution is modeled using
Chameleon profiles \citep[defined as the difference of two
non-singular $r^{-2}$ elliptical profiles;][]{kassiola1993,dutton2011}, which are
a good approximation to S\'{e}rsic profiles.  We represent the galaxy
light distribution as the sum of two Chameleon profiles with a common
centroid.  We use Chameleon profiles rather than S\'{e}rsic profiles
because they provide a similarly good fit to the data
(see Sections~\ref{subsec:sys_tests} and \ref{sec:results}) and it is more
straightforward to link their parameters to the mass parameters in our
tests of alternative mass models.  Model parameters of the lens and
source are constrained through Markov Chain Monte Carlo (MCMC)
sampling.

Since we account for G1 at a different redshift from the main lens galaxy, we make use of
the full multi-plane lens equation \citep[e.g.,][]{blandford1986,kovner1987,schneider1992,petters2001,collett2014,mccully2014} in our modeling.  We vary $H_{0}$ directly in our models and use this
distribution to calculate the effective model time-delay distance $\tdistmod$.  In calculating $\tdistmod$, we assume $\Om = 0.3$, $\OL = 0.7$, and $w = -1$, although we show that relaxing these assumptions shifts the resulting $\tdistmod$ distributions by $< 1$\% (Section~\ref{subsec:cosmo}).

\subsection{Mass Model} \label{subsec:massmodel}
Our primary mass model for the lens galaxy is a singular power-law
elliptical mass distribution \citep[hereafter ``SPEMD";][]{barkana1998}, although we also
test a model consisting of a baryonic component that traces the light
distribution and a separate dark matter component (hereafter the
``composite" model; see Section~\ref{subsec:sys_tests}).  We also
include an external shear in the strong lens plane.  Past studies have shown that a power-law model provides an good general
description of typical lens galaxies at the length scales we are interested in \citep[e.g.,][]{koopmans2006,koopmans2009,suyu2009,auger2010,barnabe2011,sonnenfeld2013}.

We also explicitly include the most nearby massive perturbing
galaxy (G1 in Figure~\ref{fig:field}; $z = 0.7821$, $\mathrm{log(M_{*}/M_{\odot})}=10.9$) that is
projected $\sim4\farcs5$ away from the lens, which is close enough
that its influence may not be adequately described by external shear
\citep[H0LiCOW Paper II; see also][]{mccully2016}.  G1 is modeled as a singular
isothermal sphere, which is a reasonable assumption as higher-order moments of its potential will have a small influence at the position of the main lens galaxy.
G1 is treated using the full multi-plane lens equation, as detailed by Suyu et al. (in preparation).

Our SPEMD model has the following free parameters:
\begin{itemize}
\item position ($\theta_{1}$,$\theta_{2}$) of the centroid (allowed to vary independently from the centroid of the light distribution)
\item Einstein radius $\theta_{\mathrm{E}}$
\item minor-to-major axis ratio $q$ and associated position angle $\theta_{q}$
\item 3-dimensional slope of the power-law mass distribution $\gamma^{\prime}$
\item external shear $\gext$ and associated position angle $\theta_{\gamma}$\footnote{$\theta_{\gamma}$ is defined to be the direction of the shear itself, i.e. orthogonal to the direction of the mass producing the shear.}
\item Einstein radius of G1
\item the cosmological parameter $H_{0}$
\end{itemize}
In principle, our lens is drawn from a selection function and the choice of model priors may introduce a bias on the inferred time-delay distance.  However, since the selection function is not well known and these biases are negligibly small for an analysis like ours \citep{collett2016}, we conservatively assume uniform priors on the model parameters.

To get a starting point for our model, we run a preliminary model
where only the positions and time delays of the lensed quasar images
are used as constraints and G1 is not included.
This preliminary model is fast and easy to optimize, and we use the
output parameters as the initial parameters of our primary model.

Our constraints on the primary lens model include the positions of the
lensed quasar images, the measured time delays, and the surface
brightness of the pixels in the ACS/F555W, ACS/F814W, and WFC3/F160W images that are fit simultaneously.
We first model the lens system individually in each band to iteratively
update the PSFs using the lensed AGN images themselves in a manner
similar to \citet{chen2016}, but with the PSF corrections and source
intensity reconstructed simultaneously in our case rather than separately (Suyu et
al. in preparation).  We then fix these ``corrected" PSFs and use them in our final
models that simultaneously use the
surface brightness distribution in all three bands as constraints.  We do not enforce any similarity of pixel values at the same spatial position across different bands.  In our MCMC sampling, we
vary the light parameters of the lens galaxy and quasar
images, the mass parameters of the lens galaxy, the external shear,
the Einstein radius of G1, and $H_{0}$.  The quasar
positions are linked across all three bands, but the other light
parameters are allowed to vary independently.

\begin{figure*}
\includegraphics[width=\textwidth]{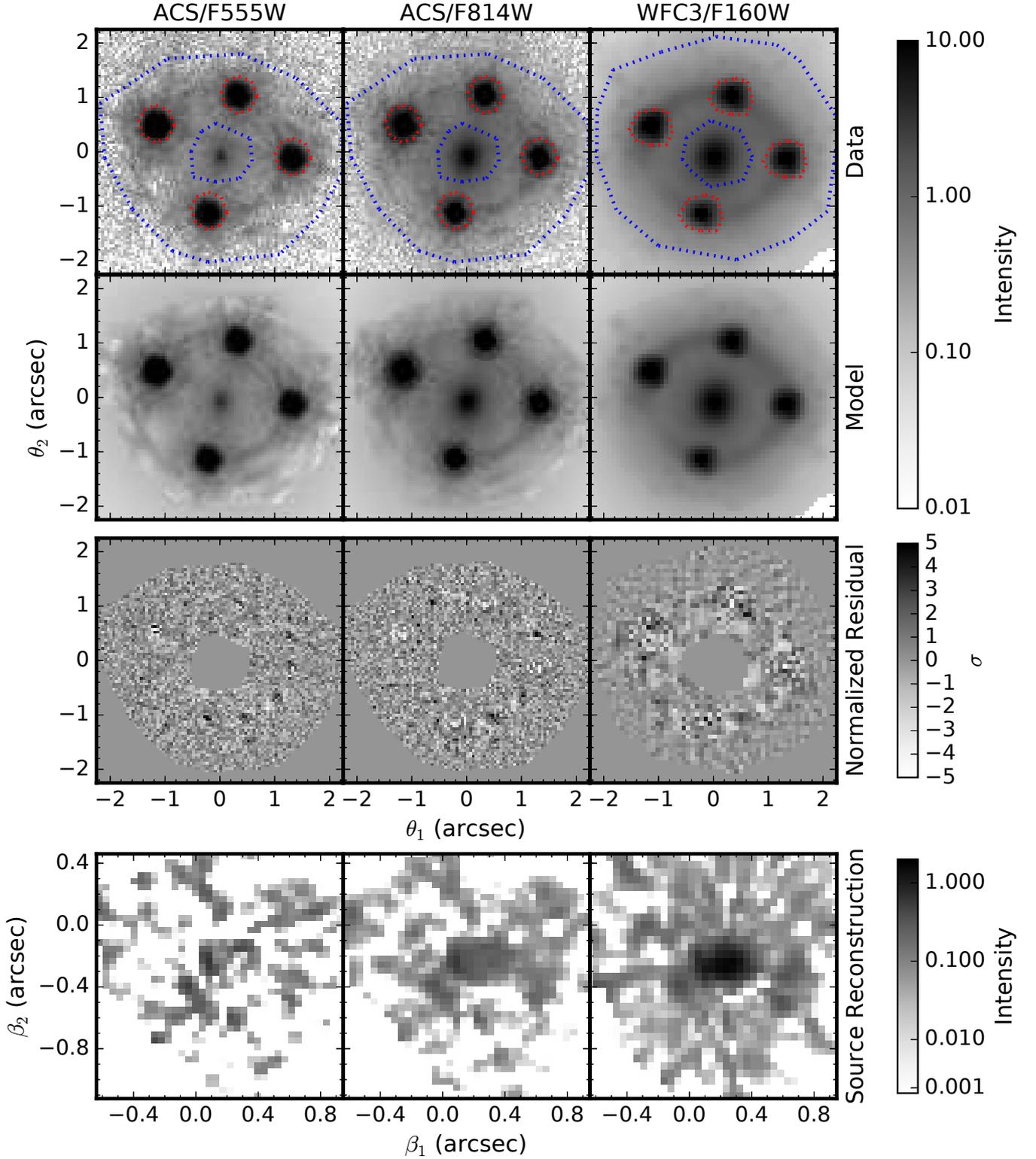}
\caption{Lens model results for ACS/F555W (left), ACS/F814W (middle),
  and WFC3/F160W (right).  Shown are the observed image (top row), the
  reconstructed image predicted by the model (second row), the
  normalized residual within the arcmask region (defined as the difference between the data and model, normalized by the estimated uncertainty of each pixel; third row), and the reconstructed source (bottom
  row).  The blue dotted lines indicate the arcmask region used for
  fitting the extended source, and the red dotted lines indicate the AGN mask region where the power-law weighting is applied.  The color bars show the scale in the
  respective panels.  The results shown here are for the fiducial model, but the results for the other systematics tests (Section~\ref{subsec:sys_tests}) are qualitatively similar.
\label{fig:model_results}}
\end{figure*}

Figure~\ref{fig:model_results} shows the data and the lens model results in each of
the three bands, as well as the source reconstruction.  Our model reproduces the surface brightness structure of the lensed AGN and host galaxy in all three bands.  There are some small residuals in the region of the lensed arc away from the AGN images.  We attribute these to compact star-formation regions in the host galaxy, as our model maps these features to similar locations in the source plane.  We test a model where the region near these residuals are masked out and find that our $\tdist$ inference is consistent to within our systematic uncertainties (Section~\ref{subsec:sys_tests}).

\subsection{Kinematics} \label{subsec:kinematics}
We follow \citet{suyu2010b} and \citet{sonnenfeld2012} to compute
the LOS stellar velocity dispersion of the strong lens galaxy
through the spherical Jeans equation \citep[see also][]{treu2002,koopmans2003}.
For a given lens model we obtain the 3D density profile of the lens galaxy by taking the spherical deprojection of the circularized surface mass density profile. The resulting 3D density profile assumes an analytical form for both the power-law and the composite model.
The 3D distribution of tracers is obtained by applying the same procedure to the surface brightness distribution of the lens galaxy, which we model as a \citet{hernquist1990} profile.  We also tested a \citet{jaffe1983} profile that has been shown to produce similar results \citep{suyu2010b}, and find that the results are affected by less than 1\% level.
We parametrize the orbital anisotropy profile with an Osipkov-Merritt model \citep{osipkov1979,merritt1985}
\begin{equation}
\frac{\sigma_\theta^2}{\sigma_r^2} = 1 - \frac{r^2}{\rani^2 + r^2}.
\end{equation}
Given values of the lens mass parameters in \sref{subsec:massmodel}, the external convergence $\kext$ in \sref{subsec:convergence} and the anisotropy radius $\rani$, we then calculate the LOS velocity dispersion profile by numerically integrating the solutions of the spherical Jeans equation as given by \citet{mamonlokas2005}.
Finally, we calculate the integral over the spectroscopic slit of the seeing-convolved brightness-weighted LOS velocity dispersion $\sigma^{\rm P}$ \citep[Equation (20) of][]{suyu2010b}, which we then compare to the measurements to compute the likelihood of the kinematics data,
\be
\label{eq:kinlike}
P(\sigma|\boldsymbol{\nu},\bm{\pi},\kext,\rani) = \frac{1}{\sqrt{2\pi}\sigma_{\sigma}} \exp\left[ -\frac{(\sigma^{\rm P}(\boldsymbol{\nu},\bm{\pi},\kext,\rani) -
    \sigma )^2}{2\sigma_{\sigma}^2}\right],
\ee
where $\sigma =222\,\kms$ and $\sigma_{\sigma}=15\,\kms$
(\sref{subsec:vddata}). We adopt a uniform prior
on $\rani$ between 0.5 and 5 times the effective radius, $\reff$,
which we determine to be $\reff=1\farcs33$ from our lens light fitting\footnote{We use the double S\'ersic model of the lens galaxy light to determine $\reff$ because the Chameleon profile does not provide an accurate description at large radii.} in the F814W filter.

\subsection{External Convergence} \label{subsec:convergence}
In H0LiCOW Paper III, we estimate the external convergence using weighted number counts in
a manner similar to \citet{greene2013} \citep[see also][]{fassnacht2011}.
We use the weighted counts to select corresponding lines of sight from the $\kext$ catalogs produced from the Millennium Simulation by \citet{hilbert2009} and, thus, to get a $\kext$ distribution.  
We use the $\kext$ distribution from H0LiCOW Paper III that was derived by combining three constraints: the unweighted galaxy number counts, the counts weighted by 1/r, and external shear matching that from our lens modeling, which gives a median external convergence at the position of \hequad~of $\kext=\kextmed$, with 16\% and 84\% percentiles of $\kext=\kextlo$ and $\kext=\kexthi$, respectively.
Although the external shear can change slightly among different models, 
these changes generally affect the $\kext$ distribution at the $\sim0.005$ level or smaller, which we can safely neglect.  Since we are explicitly including the nearest LOS perturber in our mass model, this galaxy does not contribute to the
inferred external shear, nor do we want it to be double counted in the external convergence.  We therefore exclude galaxies projected within 5\arcsec~of the main lens galaxy when
calculating the relative galaxy number counts\footnote{For our model that includes the five nearest perturbers, we run a test where we calculate $\kext$ excluding a larger region.  The corresponding shift in $\kext$ affects our final $\tdist$ distribution by $\sim0.2$\% at most, so we neglect this effect.} for both the simulated and real lines of sight.

The host galaxy group likely has a small effect as the external shear is small, and an estimate of its flexion shift (H0LiCOW Paper II) indicates that it is a less significant perturber than G1.  In addition, a weak lensing analysis of the field (Tihhonova et al. in preparation) finds a conservative $3\sigma$ upper limit of $\kext=0.04$ at the lens position, further suggesting that the group does not significantly affect our analysis.  The external convergence contribution of the host galaxy group is implicitly included in our model through the procedure of H0LiCOW Paper III.

\subsection{Blind Analysis} \label{subsec:blind}
Throughout our analysis, we blind the $H_{0}$ values in our lens model and the inferred time-delay distance
values to avoid confirmation bias using a similar procedure as
\citet{suyu2013}.
This is done by subtracting the median of the parameter PDFs from the distribution when displaying plots.  This allows us to measure the precision and relative offsets of these parameter distributions and their correlation with other parameters without being able to see the absolute value.  This eliminates the tendency for experimenters to stop investigating systematic errors when they obtain an answer consistent with the ``expected" result.  After finalizing our analysis, writing our paper draft with blinded $\tdist$ distributions, and coming to a consensus among the coauthors during a collaboration telecon on 2016 June 16, we unblind the results and do not make any further changes to the models.  There is also no iteration between the lens modeling and time delay measurements (i.e., the delays are measured once and used as they are; see H0LiCOW Paper V).  Throughout this paper, we show blinded $\tdist$ distributions until Section~\ref{sec:results}, where we reveal the absolute $\tdist$ values from our inference.

\subsection{Inferring the time-delay distance}
\label{subsec:inference}
Our inference on $\tdist$ using all of the available data is
calculated as in \eref{eq:pdf} and \eref{eq:pdf_sep}.  We use
importance sampling \citep[e.g.,][]{lewis2002} to combine the velocity dispersion and external
convergence distributions with the $\tdistmod$ inferred from our lens
model.  Specifically, for each set of lens parameters
$\boldsymbol{\nu}$ from our lens mass model MCMC chain, we draw a
sample of $\kext$ from the distribution in \sref{subsec:convergence}
and a sample of $\rani$ from the uniform distribution [0.5,5]$\reff$.
With these, we can then compute the kinematics likelihood in \eref{eq:kinlike} for the joint sample $\{\boldsymbol{\nu}, \kext,
\rani\}$ and use this to weight the joint sample.  From the effective
{\it model} time-delay distance computed from our multi-plane lensing
($\tdistmod$) and the external convergence ($\kext$), we can then
compute the effective time-delay distance ($\tdist$) via \eref{eq:ddtkappa}, keeping its absolute value blinded until we finalize our analysis.  The resulting distribution of $\tdist$ encapsulates
the cosmological information from \hequad.

\subsection{Systematics Tests} \label{subsec:sys_tests}
In this section we describe a range of tests of the effects
of various systematics in our modeling.  In
addition to a basic ``fiducial" model, we perform inferences given
the following sets of assumptions:
\begin{itemize}
\item A model with the image plane cutout region in all bands
  increased by 10 pixels in both the $\theta_{1}$ and $\theta_{2}$-directions.
\item A model with the arcmask region increased by one pixel on both
  the inner and outer edges.  To compensate for the larger arcmask region, we increase
  the source plane resolution to $50\times50$ pixels in all bands.
\item A model with the arcmask region increased by two pixels on both
  the inner and outer edges.  To accommodate the larger arcmask, we
  also increase the image plane cutout region by 10 pixels in all
  bands.  To compensate for the larger arcmask region, we increase the source plane
  resolution to $50\times50$ pixels in all bands.
\item A model where the regions near the AGN images are given zero
  weight rather than being scaled by a power-law weighting.
\item A model where the regions near the AGN images scaled by the
  power-law weighting is increased by one pixel around the outer edge.
\item A model where the regions near the AGN images scaled by the
  power-law weighting is increased by two pixels around the outer edge.
\item A model where the light profile of the lens galaxy is
  represented by the sum of two S\'{e}rsic profiles rather than the
  sum of two Chameleon profiles.
\item A model including the five most significant nearby perturbers
  (shown in Figure~\ref{fig:field}) rather than just G1.  The
  relative Einstein radii of the perturbers, assumed to be singular
  isothermal spheres, are calculated from their stellar masses (H0LiCOW Paper III),
  assuming a relationship between velocity dispersion and stellar mass from \citet{bernardi2011}.
  The ratio of Einstein radii is fixed, but with a global scaling allowed to vary freely.  This is done to
  prevent the model from optimizing the perturbers' Einstein radii in
  a way that would be inconsistent with their measured redshifts and
  stellar masses.  The galaxies' stellar masses are computed assuming the
  cosmology of the Millennium Simulation
  \citep[$H_{0} = 73~\mathrm{km~s^{-1}~Mpc^{-1}}$, $\Omega_{m} = 0.25$, $\Omega_{\Lambda} = 0.75$;][]{springel2005,hilbert2009}, but we verify that for alternative cosmologies, their stellar masses change by $< 0.02$ dex, and the ratios of their Einstein radii therefore are affected by a negligible amount.
\item A ``composite" model with separate stellar and dark matter
  components.  The details of this model are discussed in
  Section~\ref{subsec:pl_comp}.
\item The composite model with the regions near the AGN images scaled by the
  power-law weighting increased by one pixel around the outer edge.
\item The composite model with the arcmask region increased by one pixel on both
  the inner and outer edges and a $50\times50$ pixel source plane resolution.
\end{itemize}

As described in Section~\ref{subsec:joint}, we combine the MCMC chains
from all of these tests.
In doing so, we effectively assume that 1) these various
tests sample a reasonable distribution of assumptions that we could have
made when modeling the system, and that these assumptions have equal
prior probability, and 2) neither the goodness of fit nor the
parameter space prior volume are appreciably different between the tests.
We verify that the goodness of fit does not change appreciably during
this procedure (see Section~\ref{sec:results}).
We weight the different MCMC chains equally and concatenate
them, resulting in a set of samples that characterizes our
final posterior PDF for $\tdist$. This procedure folds the systematic uncertainty due to our modeling
assumptions into our final uncertainty on the inferred parameters.

\subsection{Comparison of Power Law and Composite Models} \label{subsec:pl_comp}
We follow \citet{suyu2014} to construct the composite model of baryons
and dark matter as one of our systematics tests.  The composite model consists of mass components associated with each
of the four non-singular isothermal elliptical profiles (making up the
two Chameleon profiles) in the lens galaxy light model in the
WFC3/F160W band scaled by an overall mass-to-light (M/L) ratio.  We use
the F160W band because it probes the rest-frame near-infrared and thus
should be the best tracer of stellar mass.  The dark matter component
is modeled as an elliptical NFW \citep{navarro1996} potential with its
centroid linked to that of the light centroid in F160W.  This is motivated by \citet{dutton2014}, who find that non-contracted NFW profiles are a good representation for the dark matter halos of massive elliptical galaxies.

The composite model has the following free parameters:
\begin{itemize}
\item M/L ratio for the baryonic component
\item NFW halo normalization $\kappa_{0,h}$ \citep[defined as $\kappa_{0,h} \equiv 4\kappa_{s}$;][]{golse2002}
\item NFW halo scale radius $r_{s}$
\item NFW halo minor-to-major axis ratio $q$ and associated position angle $\theta_{q}$
\item external shear $\gext$ and associated position angle $\theta_{\gamma}$
\item Einstein radius of G1
\item the cosmological parameter $H_{0}$
\end{itemize}
We set a Gaussian prior of $r_{s} = 14\farcs3 \pm 2\farcs0$ based on the results of
\citet{gavazzi2007} for lenses in the Sloan Lens ACS Survey \citep[SLACS;][]{bolton2006} sample, which encompasses
the redshift and stellar mass of \hequad.
All other parameters are
given uniform priors.  We note that the relative amplitudes of the two
Chameleon profiles representing the stellar light distribution of the
lens galaxy can vary during the modeling, whilst the relative
amplitudes are fixed in the mass profiles.  To account for this, we
adopt an iterative approach where we run a series of MCMC chains and
update the (fixed) relative amplitudes of the associated mass components to
match that of the light components after each chain.  We iterate until the relative
change in the light profile amplitudes reach a point where the
inferred $\tdist$ stabilizes, then combine the MCMC chains after this
point into a single distribution to represent the composite model.  The remaining two composite models (with a larger arcmask or AGN mask) use fixed relative amplitudes of the mass components from the latest iteration of the original composite model.

\begin{figure*}
\includegraphics[width=\textwidth]{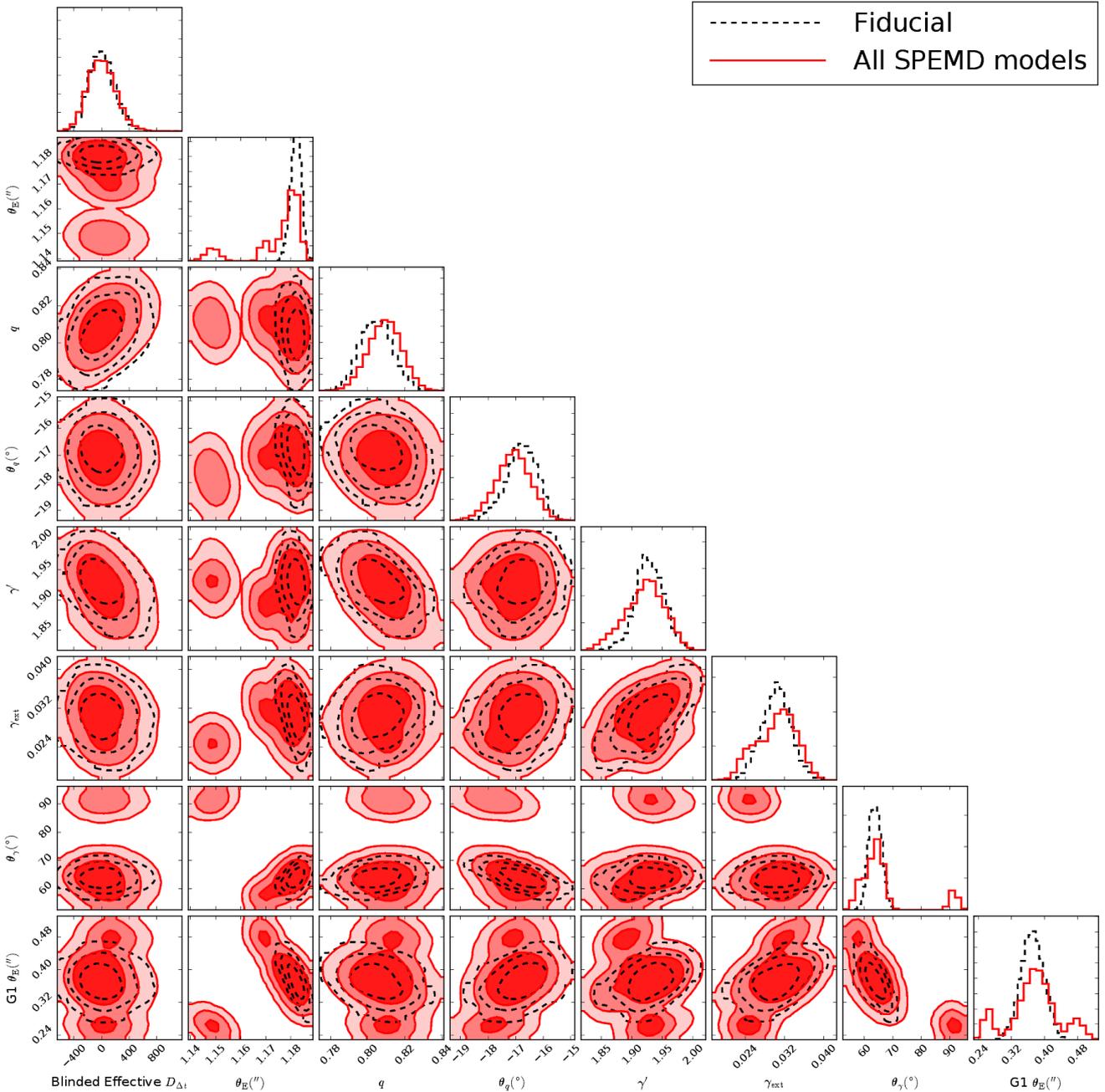}
\caption{
Marginalized parameter distributions from our SPEMD lens model
results.  We show the fiducial model (dashed black contours) and the combined results
from our systematics tests (shaded red contours).  The contours represent
the 68.3\%, 95.4\%, and 99.7\% quantiles.
\label{fig:corner_spemd}}
\end{figure*}

The marginalized parameter distributions of the SPEMD model are shown in Figure~\ref{fig:corner_spemd}.  We show the combined distributions of all SPEMD models as well as the fiducial model separately.  The parameter statistics for each model are given in Appendix~\ref{app:params}.  We note that two particular models stand out.  The model with the arcmask expanded by one pixel and a $50\times50$ source grid prefers a smaller Einstein radius for the main lens galaxy and a larger Einstein radius for G1.  This degeneracy is likely due to systematics associated with the source pixel size \citep{suyu2013}, as this model has a smaller source pixel size than the others.  The 5-perturber model prefers a smaller Einstein radius for both the main lens galaxy and G1, as well as a very different $\theta_{\gamma}$.  This is not surprising, as the addition of the extra perturbers in the lens model contributes to the integrated LOS lensing effect, reducing the contribution needed from the main lens and G1, as well as changing the external shear needed to fit the data.  The offset between the mass centroid and the light centroid in the F160W band is typically $\sim0.002\arcsec$.

\begin{figure*}
\includegraphics[width=\textwidth]{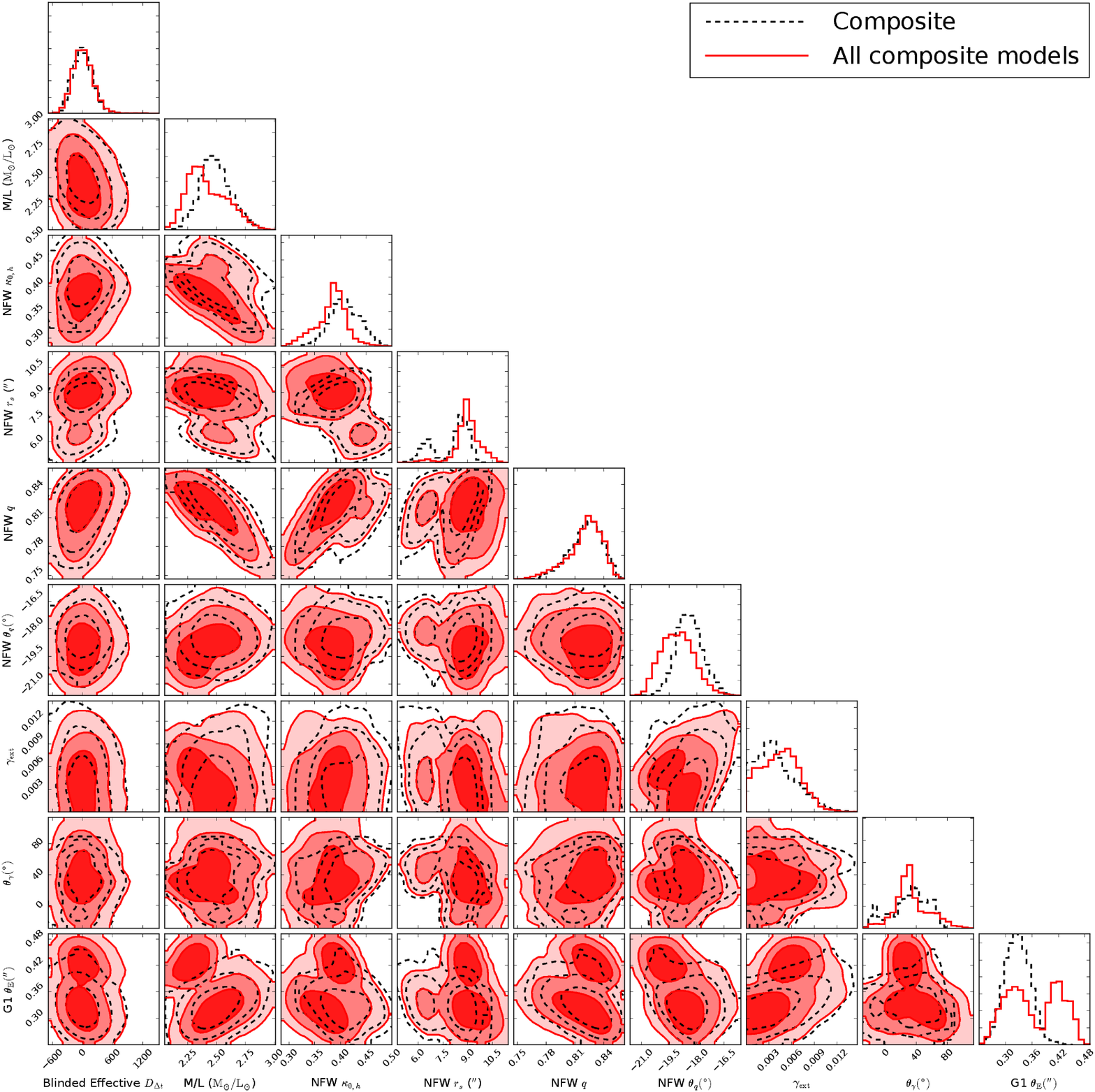}
\caption{
Marginalized parameter distributions from our composite lens model
results.  We show the main composite model (dashed black contours) and the combined results
from our systematics tests (shaded red contours).  The contours represent
the 68.3\%, 95.4\%, and 99.7\% quantiles.
\label{fig:corner_compos}}
\end{figure*}

We show the marginalized parameter distributions of the composite model in Figure~\ref{fig:corner_compos}.  Again, we show the combined distributions as well as the main composite model separately, and the parameter statistics for each model are given in Appendix~\ref{app:params}.  The main composite model appears to have some degenerate or bimodal features, but this is because this model itself is the combination of several separate models with slightly different relative amplitudes between the two Chameleon components, as mentioned above.  The model with a larger arcmask and source grid prefers a larger G1 Einstein radius, similar to the analogous SPEMD model.  The dark matter fraction within the Einstein radius for the composite models is $f_{DM} \sim 45$\%.

We compare the physical parameters of our ``fiducial" power law model
to the composite model.  The results are shown in
Table~\ref{tab:fid_comp}, with the parameter statistics for all composite models given in Appendix~\ref{app:params}.  We note that the external shear strength of the composite model is smaller than that of the power law model, which we attribute to a degeneracy between $\gext$ and the internal ellipticity of the mass model.  When external shear is removed, the composite model's critical curves appear slightly more elliptical than those of the power law model, supporting this interpretation.  As mentioned in Section~\ref{subsec:convergence}, the difference in $\gext$ between these models has a negligible effect on the $\kext$ distribution.

\renewcommand*\arraystretch{1.5}
\begin{table}
\caption{Lens Model Parameters \label{tab:fid_comp}}
\begin{minipage}{\linewidth}
\begin{tabular}{l|c}
\hline
Parameter &
Marginalized Constraints
\\
\hline
\multicolumn{2}{l}{Fiducial Singular Power Law Ellipsoid Model}
\\
\hline
$\theta_{\mathrm{E}}~(\arcsec)$\footnote{Spherical-equivalent Einstein radius} &
$1.182_{-0.002}^{+0.002}$
\\
$q$ &
$0.80_{-0.01}^{+0.01}$
\\
$\theta_{q}$ ($^{\circ}$) &
$-16.8_{-0.6}^{+0.5}$
\\
$\gamma^{\prime}$ &
$1.93_{-0.02}^{+0.02}$
\\
$\gext$ &
$0.030_{-0.003}^{+0.003}$
\\
$\theta_{\gamma}$ ($^{\circ}$) &
$63.7_{-2.2}^{+2.4}$
\\
G1 $\theta_{\mathrm{E}}~(\arcsec)$ &
$0.37_{-0.03}^{+0.03}$
\\
\hline
\multicolumn{2}{l}{Composite Model}
\\
\hline
Stellar M/L ($\mathrm{M_{\odot}/L_{\odot}}$)\footnote{M/L within $\theta_{\mathrm{E}}$ for rest-frame $V$ band.  The given uncertainties are only statistical and do not include systematic effects.  The stellar mass is calculated assuming $H_{0} = 70~\mathrm{km~s^{-1}~Mpc^{-1}}$, $\Omega_{m} = 0.3$, $\Omega_{\Lambda} = 0.7$, but changes in the cosmology affect the M/L by a negligible amount.} &
$2.5_{-0.1}^{+0.1}$
\\
NFW $\kappa_{0,h}$ &
$0.41_{-0.03}^{+0.03}$
\\
NFW $r_{s}~(\arcsec)$ &
$8.43_{-1.94}^{+0.58}$
\\
NFW $q$ &
$0.82_{-0.02}^{+0.01}$
\\
NFW $\theta_{q}$ ($^{\circ}$) &
$-18.4_{-0.7}^{+0.7}$
\\
$\gext$ &
$0.004_{-0.002}^{+0.003}$
\\
$\theta_{\gamma}$ ($^{\circ}$) &
$34.4_{-32.5}^{+22.9}$
\\
G1 $\theta_{\mathrm{E}}~(\arcsec)$ &
$0.33_{-0.03}^{+0.03}$
\\
\hline
\end{tabular}
\\
{\footnotesize Reported values are medians, with errors corresponding to the 16th and 84th percentiles.}
\\
{\footnotesize Angles are measured east of north.}
\end{minipage}
\end{table}
\renewcommand*\arraystretch{1.0}

\subsection{Impact of Different Cosmologies}
\label{subsec:cosmo}
\begin{figure}
\includegraphics[width=0.5\textwidth]{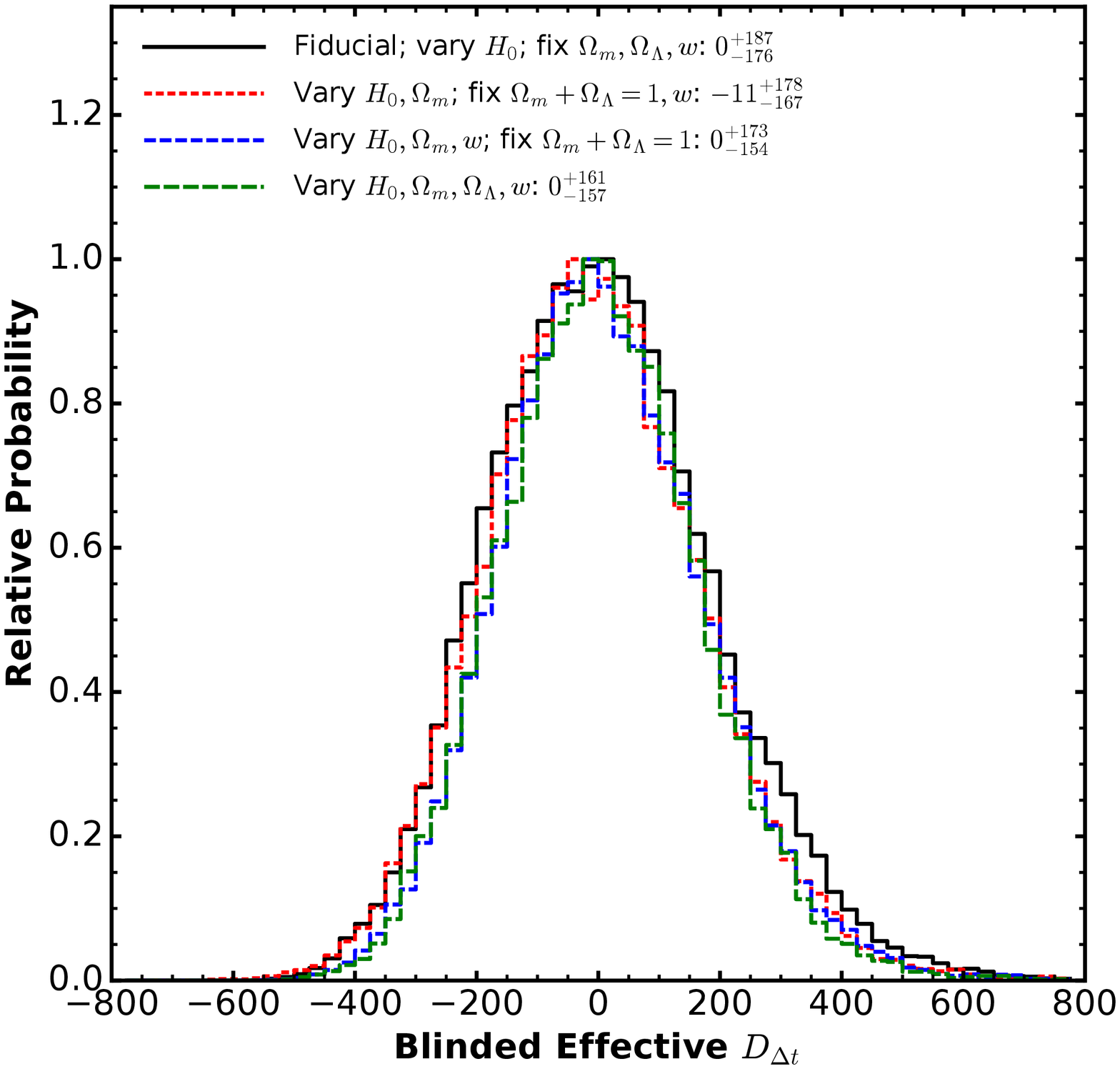}
\caption{
PDF of $\tdist$ for the various cosmologies.  We compare the fiducial model to one in which $\Om$ is allowed to vary (with $\Om + \OL = 1$), one in which $w$ is also allowed to vary, and one in which $\Om$, $\OL$, and $w$ are all allowed to vary independently.  The distributions are blinded by subtracting the median of the fiducial model PDF.  The different cosmology tests are indicated by the legend, and the median and 68\% quantiles of the $\tdist$ distributions are given.  The median of the blinded effective time-delay distance PSF is insensitive to the extra cosmological parameters to within 1\%.
\label{fig:dt_cosmo}}
\end{figure}

In the multi-lens-plane modeling, we need to sample the cosmological
parameters in order to carry out the ray tracing.
Throughout our analysis, we only vary $H_{0}$, keeping other
cosmological parameters fixed ($\Om = 0.3$, $\OL =
0.7$, $w = -1$).  This is done for computational reasons, as the MCMC sampling becomes inefficient when they are all allowed to vary simultaneously.  In principle, $\tdist$ has a weak dependence
on these other cosmological parameters.  We test their impact by rerunning the fiducial model while allowing combinations of them to vary with uniform priors.  The resulting effective $\tdist$ distributions, shown in
Figure~\ref{fig:dt_cosmo}, have peaks that are consistent to within $1\%$ of the absolute value, demonstrating that the results are
insensitive to these extra cosmological parameters at the level of accuracy that we are currently working at.  In future, when errors shrink further, this sampling will be included.
We conclude that with the current level of precision, we are justified in deriving the posterior distribution function of the time-delay distance by varying $H_{0}$ only for computational efficiency.  We emphasize that this does not affect in any way the generality of our results and that the resulting posterior distribution function is robust and can be interpreted in any cosmological model.

To expand on this point, it is instructive to consider multi-plane
lensing \citep[e.g.,][]{blandford1986,kovner1987,kochanek1988,schneider1992,petters2001,collett2014,mccully2014,schneider2014a} for the case of two
lens planes, as we have in most of our models.  Defining $\bm{\theta}_{1}$, $\bm{\theta}_{2}$, and
$\bm{\theta}_3$ as the angular coordinates on the main lens plane, the
G1 lens plane, and the source plane, respectively, the multi-plane lens
equations in this case are
\bea
\label{eq:mp1a}
\bm{\theta}_{2} = \bm{\theta}_{1} - \frac{D_{12}}{D_2}
\bm{\hat{\alpha}}_{1}(D_1 \bm{\theta}_{1}),
\\
\label{eq:mp1b}
\bm{\theta}_{3} = \bm{\theta}_{1} - \frac{D_{13}}{D_3}
\bm{\hat{\alpha}}_{1}(D_1 \bm{\theta}_{1}) - \frac{D_{23}}{D_3}
\bm{\hat{\alpha}}_{2}(D_2 \bm{\theta}_{2}),
\eea
where $D_i$ is the angular diameter distance from the observer to plane $i$,
$D_{ij}$ is the angular diameter distance between planes $i$ and $j$,
and $\bm{\hat{\alpha}}_{i}$ is the deflection angle at plane $i$.
Scaling the deflection angle relative to the source (third) plane, we
have the scaled deflection angle as
\be
\label{eq:salpha}
\bm{\alpha}_i(\bm{\theta}_{i}) = \frac{D_{i3}}{D_3}
\bm{\hat{\alpha}}_i(D_i \bm{\theta}_{i}).
\ee
By further defining
\be
\label{eq:beta}
\beta_{ij} = \frac{D_{ij}}{D_j}\frac{D_3}{D_{i3}},
\ee
we can rewrite Equations (\ref{eq:mp1a}) and (\ref{eq:mp1b}) as
\bea
\label{eq:mp2a}
\bm{\theta}_{2} = \bm{\theta}_{1} - \beta_{12} \bm{\alpha}_{1}(\bm{\theta}_{1}),
\\
\label{eq:mp2b}
\bm{\theta}_{3} = \bm{\theta}_{1} - \bm{\alpha}_{1}(\bm{\theta}_{1}) -
\bm{\alpha}_{2}(\bm{\theta}_{2}).
\eea
The multi-plane time delay has contributions from the geometric delays
between planes and the gravitational delay at each mass plane:
\bea
\nonumber
t = & \frac{\tdist^{12}}{c}  \left[ \frac{1}{2}
  |\bm{\theta}_{2} - \bm{\theta}_{1}|^{2} - \beta_{12}
  \psi_{1}(\bm{\theta}_{1}) \right] \\
\label{eq:mptd}
   & + \frac{\tdist^{23}}{c} \left[ \frac{1}{2} |\bm{\theta}_{3} -
  \bm{\theta}_{2}|^{2} - \psi_{2}(\bm{\theta}_{2}) \right],
\eea
where $\psi_{i}$ is the lens
potential related to the scaled deflection angle via $\nabla
\psi_{i}=\bm{\alpha}_i$, and the time-delay distances between planes
are
\be
\tdist^{ij} \equiv (1+z_i) \frac{D_i D_j}{D_{ij}},
\ee
with $z_i$ being the redshift of plane $i$.  From \eref{eq:mptd}, we
see that the time delay depends on the two time-delay distances and
$\beta_{12}$.  In general it is difficult to constrain all these distance
quantities independently.  In
fact, in multi-plane modeling, we adopt specific cosmological models
to compute the distances ($D_{ij}$ and ${D_i}$) for the ray tracing,
and compare the
time-delay distance measurements from these different background
cosmologies.  For the case of \hequad~where G1 is not strongly
lensing the background source but merely perturbs it, the effect on
the time delays from G1 is weak.  The
lack of sensitivity to $\Om$ and $w$ suggests that \hequad~is not
sensitive to $\beta_{12}$ at an interesting level to probe it
directly in the same way as a double source plane lens
\citep[e.g.,][]{gavazzi2008, collett2014}.  In \hequad, we find
instead that the time delays are mostly set by the strong lens, and we
can measure the effective $\tdist$, which is $\tdist^{13}$, that is
independent of assumptions on the background cosmology, as
demonstrated in \fref{fig:dt_cosmo}.  This robust distance
determination then permits
us to constrain any reasonable cosmological model via the distance-redshift
relation.

%-------------------------------------------------------------------------------

\section{Results} \label{sec:results}
\begin{figure*}
\includegraphics[width=\textwidth]{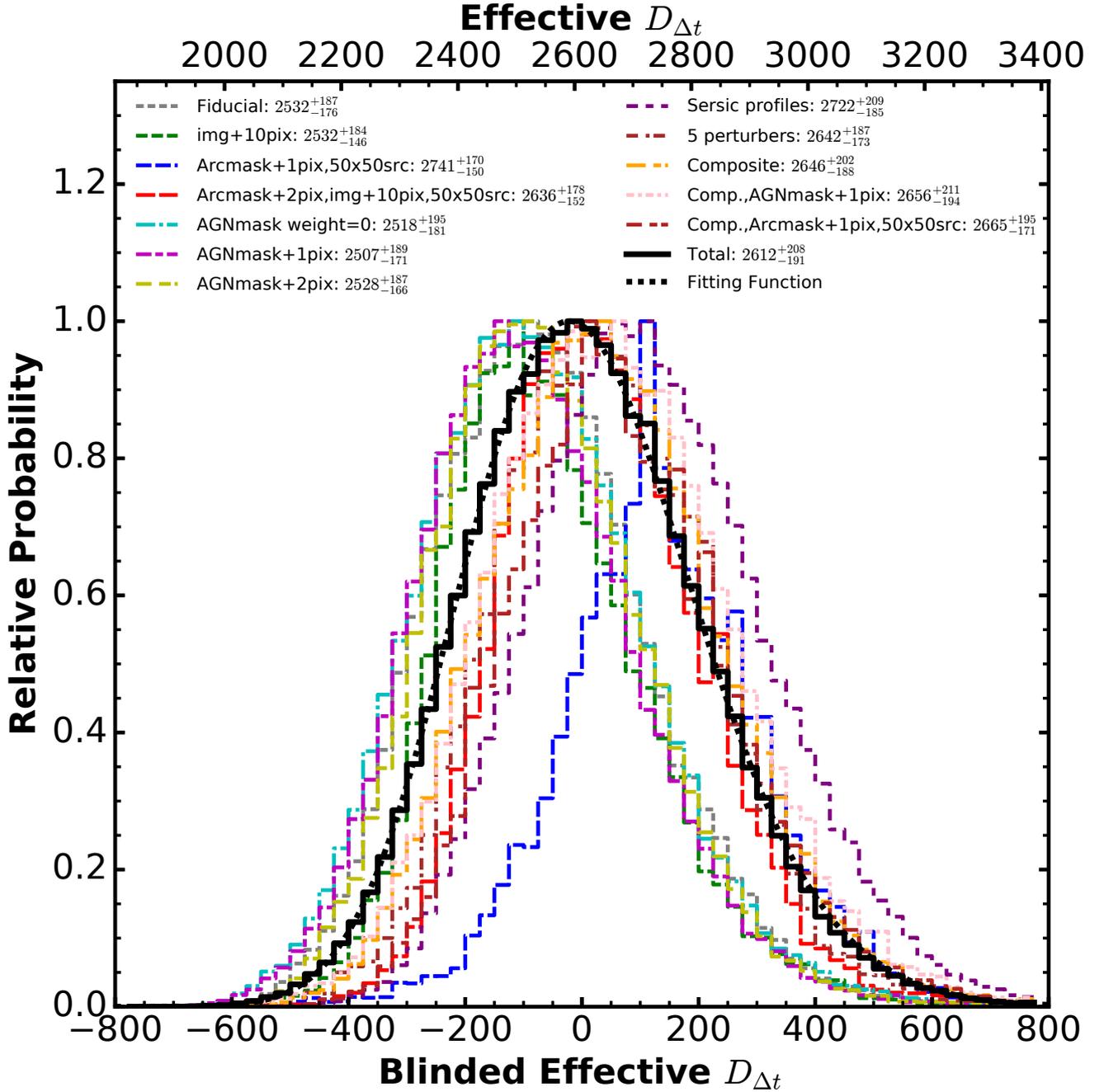}
\caption{
PDF of $\tdist$ for the various models, as
indicated by the legend.  The median and 68\% quantile of each distribution is given.  The thick black line represents the sum of
all the distributions, which accounts for the various systematic uncertainties.  The dotted black line is the skewed lognormal
distribution (\eref{eq:pdt}) fit to the final distribution.  The bottom x-axis shows the blinded result, which is obtained by subtracting the median of the combined PDF from the absolute $\tdist$ values.  The top x-axis shows the true $\tdist$ values.  Throughout our blind analysis, the top x-axis was hidden until our analysis was finalized.
\label{fig:dt_sys}}
\end{figure*}

The marginalized posterior $\tdist$ distributions for our lens
model are given in Table~\ref{tab:dt}.  We report the median and 68\% quantiles for each of the models described in
Section~\ref{subsec:sys_tests}, as well as a final distribution that
combines all of the chains.  These distributions are shown in
Figure~\ref{fig:dt_sys}.  The blinded distributions, shown on the bottom x-axis of Figure~\ref{fig:dt_sys}, were the only values seen until the unblinding.  The velocity dispersion and
external convergence have been included in these distributions.  Each of the chains representing a different
systematics test is given equal weight because the goodness-of-fit is comparable.  Our final constraint on the effective time-delay distance in \hequad~is $\tdist=\dt$.  We note that our fiducial model parameters are consistent with an identical model run only using the F160W band as constraints.

\renewcommand*\arraystretch{1.5}
\begin{table}
\caption{Effective Time-Delay Distance \label{tab:dt}}
\begin{minipage}{\linewidth}
\begin{tabular}{l|cc}
\hline
Model &
$D_{\Delta t}$ (Mpc) &
$\chi^{2}$
\\
\hline
Fiducial &
$2532_{-176}^{+187}$ &11024.9
\\
S\'{e}rsic profiles &
$2722_{-185}^{+209}$ &11001.5
\\
5 perturbers &
$2642_{-173}^{+187}$ &11002.0
\\
Composite &
$2646_{-188}^{+202}$ &11014.1
\\
\hline
AGN mask+1pix &
$2507_{-171}^{+189}$ &11029.6
\\
Composite,AGN mask+1pix &
$2656_{-194}^{+211}$ &11032.2
\\
\hline
Arcmask+1pix,50x50src &
$2741_{-150}^{+170}$ &11097.7
\\
Composite,Arcmask+1pix,50x50src &
$2665_{-171}^{+195}$ &11121.5
\\
\hline
img+10pix &
$2532_{-146}^{+184}$ &11090.2
\\
\hline
Arcmask+2pix,img+10pix,50x50src &
$2636_{-152}^{+178}$ &11074.3
\\
\hline
AGN mask weight=0 &
$2518_{-181}^{+195}$ &10921.6
\\
\hline
AGN mask+2pix &
$2528_{-166}^{+187}$ &11065.4
\\
\hline
{\bf Total} &
$\bm{2612_{-191}^{+208}}$ &--
\\
\hline
\end{tabular}
\\
{\footnotesize Reported values are medians, with errors corresponding to the 16th and 84th percentiles.}
\\
{\footnotesize $\chi^{2}$ values are computed within the fiducial arcmask and outside the AGNmask+2pix region for a fair comparison among models.  The models are grouped such that those that use the same dataset are together.}
\\
{\footnotesize For models with a larger arcmask, we calculate $\chi^{2}$ for a source grid resolution that approximately matches that of the other models so that we can fairly compare them.}
\end{minipage}
\end{table}
\renewcommand*\arraystretch{1.0}

Table~\ref{tab:dt} also shows the $\chi^{2}$
for each model.  The $\chi^{2}$ values are calculated within the fiducial arcmask and outside of the AGN mask+2 pixel region to ensure a fair comparison among the different models.  The $\chi^{2}$ is calculated by summing the square of the normalized residual pixels (third row of Figure~\ref{fig:model_results}) within this region.  The number of degrees of freedom, $\mathrm{N_{dof}}$, is the number of pixels in this region across all three bands ($\mathrm{N_{d}}=9577$) minus the number of lens mass/light model parameters minus a $\Gamma$ term that represents the effective number of source pixels accounting for source regularization \citep[see][]{suyu2006}.  $\Gamma$ is calculated separately for each of the models' arcmask and AGN mask regions.  The typical $\mathrm{N_{dof}}$ for our models is $\sim8400-8600$.  Most of the residual $\chi^{2}$ is associated with a few compact star-forming regions in the host galaxy that cannot be modeled at the resolution of our source pixel grid (Figure~\ref{fig:model_results}).  Our tests show that masking out these regions affects the $\tdist$ distribution by less than our systematic uncertainties (see Section~\ref{subsec:massmodel}).
We note that for a fair comparison, the $\chi^{2}$
for models with larger arcmasks are calculated on a source plane pixel scale that gives them approximately the same source resolution as the other models ($41\times41$ pixels for the arcmask+1 pixel models, $45\times45$ pixels for the arcmask+2 pixel model).
The typical absolute change in $\chi^{2}$ for one-pixel changes\footnote{A one-pixel change in source grid resolution roughly corresponds to the changes in source pixel size across our different models.} in the source grid resolution is $\sim60-70$.  We take this as the uncertainty in $\chi^{2}$, and the $\chi^{2}$ values are all very close among models that use the same dataset.  Therefore,
we are justified in weighting each of the models equally.

We fit a skewed lognormal function to the $\tdist$ distribution, as this function provides an accurate parameterized representation of our result \citep{suyu2010b}.  The distribution has the
form
\begin{equation}\label{eq:pdt}
P(\tdist) = \frac{1}{\sqrt{2\pi}(x-\lambda_{D})\sigma_{D}}~\mathrm{exp} \left[ - \frac{(\mathrm{ln}(x-\lambda_{D})-\mu_{D})^{2}}{2\sigma_{D}^{2}} \right],
\end{equation}
where $x=\tdist / (1~\mathrm{Mpc})$, $\lambda_{D} = 653.9$,
$\mu_{D} = 7.5793$, and $\sigma_{D} = 0.10312$.  We plot this best-fitting function
along with the final $\tdist$ distribution in
Figure~\ref{fig:dt_sys}.  The median, 68\%, and 95\% quantiles of the
$\tdist$ distribution and the best-fitting function agree to within $\sim0.1\%$, indicating that this function is an accurate representation.

\begin{figure}
\includegraphics[width=0.5\textwidth]{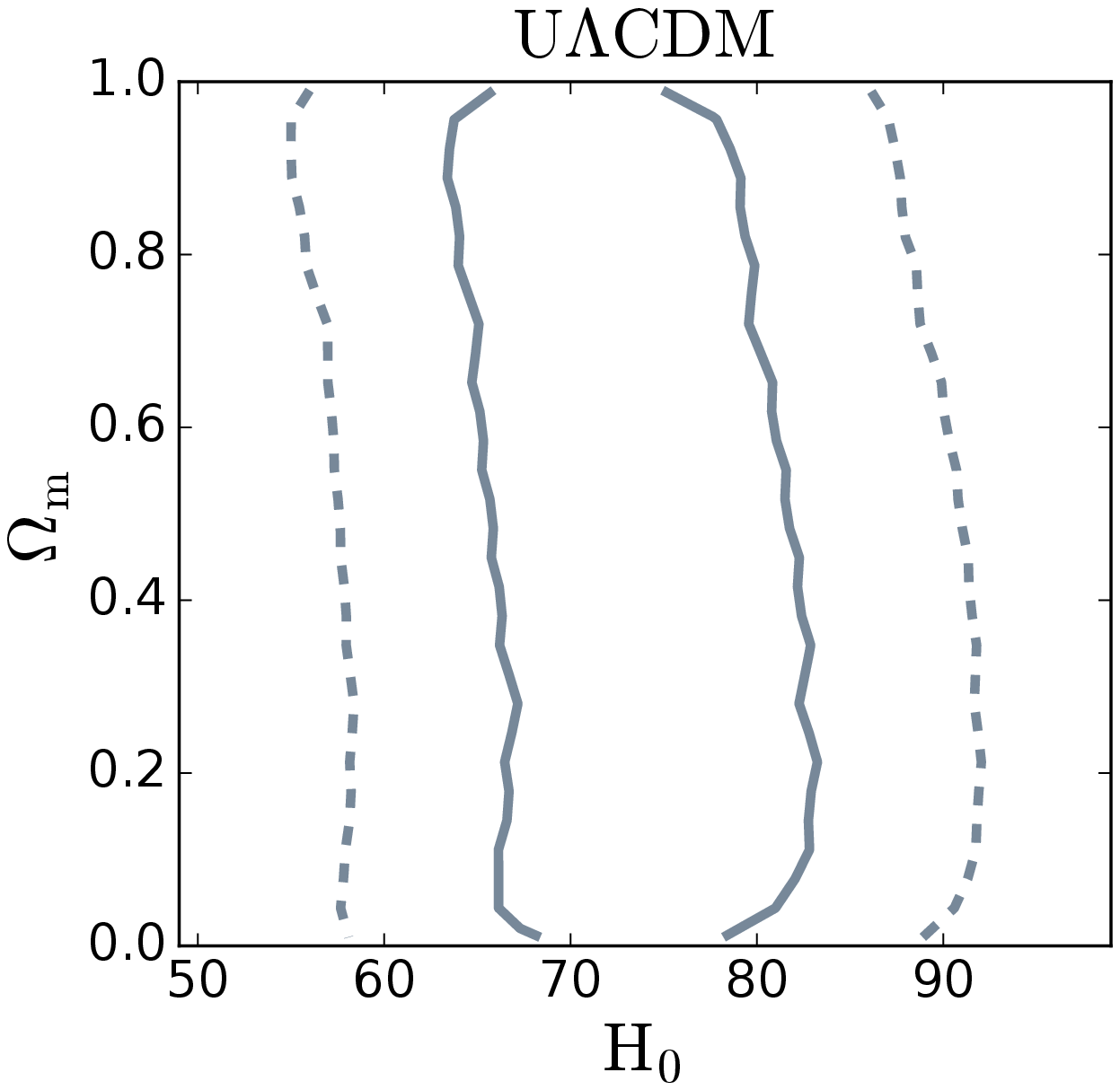}
\caption{
Posterior distribution of $H_{0}$ and $\Om$ for the U$\Lambda$CDM cosmology determined from the time-delay distance inference of \hequad.  The contours represent the 68\% and 95\% quantiles of the distribution.  $\Om$ has a weak influence on $\tdist$, so it is not well-constrained.  The marginalized value of $H_{0}$ for this cosmology is $\ulcdm$.
\label{fig:ulcdm}}
\end{figure}

Based on our inferred effective time-delay distance, we can calculate cosmological parameters for a variety of cosmological models, which are described in Table~\ref{tab:cosmo}.
For the U$\Lambda$CDM cosmology, we constrain the Hubble constant to be $H_{0} = \ulcdm$, giving a precision of $\sim8$\% from just this single lens system.  This value is in good agreement with the latest distance ladder results \citep[$H_{0} = 73.24 \pm 1.74~\mathrm{km~s^{-1}~Mpc^{-1}}$;][]{riess2016} and higher than the latest {\it Planck} measurement for a similar cosmology \citep[$H_{0} = 67.8 \pm 0.9~\mathrm{km~s^{-1}~Mpc^{-1}}$;][]{planck2015}.
Figure~\ref{fig:ulcdm} shows the posterior distribution of $H_{0}$ and $\Om$ in U$\Lambda$CDM.  Fixing $\OL$ in the UH$_{0}$ model does not change the inferred $H_{0}$ significantly ($H_{0} = \uh$).  Our results for the o$\Lambda$CDM $+$ {\it Planck} model suggest a Universe consistent with spatial flatness.  Interestingly, the wCDM $+$ {\it Planck} model prefers a dark energy equation of state parameter that is in mild tension with $w = -1$ at the $\sim2\sigma$ level.  The results for each of our models are summarized in Table~\ref{tab:cosmo}.

\renewcommand*\arraystretch{1.5}
\begin{table*}
\caption{Cosmological Parameter Constraints from \hequad \label{tab:cosmo}}
\begin{minipage}{\linewidth}
\begin{tabular}{c|ccc}
\hline
Model Name &
Description &
Parameter Priors &
Marginalized Cosmological Parameters
\\
\hline
\multirow{2}{*}{UH$_{0}$} &
Flat $\Lambda$CDM cosmology, fixed $\OL$ &
\multirow{2}{*}{$H_{0}$ uniform in [0,150]} &
\multirow{2}{*}{$H_{0}=74.3_{-5.4}^{+6.0}$}
\\
 &
$\Om = 1-\OL = 0.32$ &
 &
\\
\hline
\multirow{2}{*}{U$\Lambda$CDM} &
Flat $\Lambda$CDM cosmology &
$H_{0}$ uniform in [0,150] &
$H_{0} = 73.1_{-6.0}^{+5.7}$
\\
 &
$\Om = 1-\Omega_{\Lambda}$ &
$\OL$ uniform in [0,1] &
$\OL = 0.51_{-0.34}^{+0.34}$
\\
\hline
\multirow{4}{*}{o$\Lambda$CDM + {\it Planck}} &
 &
\multirow{4}{*}{{\it Planck} prior for \{$H_{0}$, $\OL$, $\Om$\}} &
$H_{0} = 63.5_{-3.7}^{+3.7}$
\\
 &
 Non-flat $\Lambda$CDM cosmology &
 &
$\Om = 0.35_{-0.04}^{+0.04}$
\\
 &
 $\Ok = 1 - \OL - \Om$ &
 &
$\OL = 0.66_{-0.03}^{+0.03}$
\\
 &
 &
 &
$\Ok = -0.011_{-0.011}^{+0.010}$
\\
\hline
\multirow{3}{*}{wCDM + {\it Planck}} &
\multirow{2}{*}{Flat wCDM cosmology} &
 &
$H_{0} = 83.7_{-9.0}^{+9.2}$
\\
 &
\multirow{2}{*}{$\Om = 1-\OL$} &
{\it Planck} prior for \{$H_{0}$, $\OL$, $w$\} &
$\OL = 0.80_{-0.05}^{+0.04}$
\\
 &
 &
 &
$w = -1.52_{-0.27}^{+0.27}$
\\
\hline
\end{tabular}
\\
{\footnotesize Reported values are medians, with errors corresponding to the 16th and 84th percentiles.}

{\footnotesize {\it Planck} priors are the \citet{planck2015} chains from baseline high-L {\it Planck} power spectra and low-L temperature and LFI polarization ({\it plikHM\_TT\_lowTEB}).}
\end{minipage}
\end{table*}
\renewcommand*\arraystretch{1.0}

The results for \hequad~presented here can be combined with previous analyses of \blens~\citep{suyu2010b} and \rxjlens~\citep{suyu2013,suyu2014} to produce stronger constraints on cosmology.  A full analysis of the implications of our $\tdist$ inference for a variety of cosmologies using constraints from all three H0LiCOW lenses analyzed to date is presented in H0LiCOW Paper V.

%-------------------------------------------------------------------------------

\section{Conclusions} \label{sec:conclusions}
We have performed a blind analysis of the gravitational lens \hequad~ using new deep
\hst~imaging, high-precision time delays from COSMOGRAIL, a measurement of the lens galaxy velocity dispersion, and spectroscopic and photometric data to constrain the mass distribution
along the line of sight.  Our model is able to reproduce the surface brightness structure of the lensed AGN and host galaxy in all three ${\it HST}$ bands, as well as the measured time delays.  Combining these datasets and accounting for various sources of systematic uncertainty in the lens modeling, we constrain the effective time-delay distance to be $\tdist = \dt$, giving a precision of $\dtprec\%$.  For a flat $\Lambda$CDM cosmology with uniform priors on $H_{0}$ and $\OL$, we constrain the Hubble constant to be $H_{0} = \ulcdm$ (a precision of $\sim8$\%), in good agreement with the latest distance ladder results.  A detailed analysis of the implications of our $\tdist$ constraint on a variety of cosmologies is presented in H0LiCOW Paper V.

Upcoming analyses of the remaining two H0LiCOW systems will complete the sample of five time-delay lenses and constrain $H_{0}$ to $< \combprec\%$ precision.  Our extensive blind analysis of \hequad~demonstrates the utility of gravitational lens time delays as a precise and independent cosmological probe. With hundreds of new lensed AGN expected to be discovered in current and future wide-field imaging surveys \citep{oguri2010}, we expect time-delay cosmography to provide competitive cosmological constraints throughout the next decade.

%-------------------------------------------------------------------------------

\section*{Acknowledgements}
We thank the referee for reviewing this paper and providing helpful commentary.
We thank Adriano Agnello, Roger Blandford, Geoff Chih-Fan Chen, Xuheng Ding,
Yashar Hezaveh, Kai Liao, John McKean, Georges Meylan,
Danka Paraficz, Chiara Spiniello, Malte Tewes, Olga Tihhonova, and
Simona Vegetti for their contributions to the H0LiCOW project.  We
thank Simon Birrer for useful discussions and feedback.  We thank
Bau-Ching Hsieh for computing support on the SuMIRe computing cluster.
We thank Michelle Wilson for sharing the lens galaxy group properties from an ongoing independent study.
H0LiCOW and COSMOGRAIL are made possible thanks to the continuous work of all observers and technical staff obtaining the monitoring observations, in particular at the Swiss Euler telescope at La Silla Observatory. Euler is supported by the Swiss National Science Foundation.
K.C.W. is supported by an EACOA Fellowship awarded by the East Asia Core
Observatories Association, which consists of the Academia Sinica
Institute of Astronomy and Astrophysics, the National Astronomical
Observatory of Japan, the National Astronomical Observatories of the
Chinese Academy of Sciences, and the Korea Astronomy and Space Science
Institute.
S.H.S.~is supported by the Max Planck Society through the Max
Planck Research Group.  This work is supported in part by the Ministry
of Science and Technology in Taiwan via grant
MOST-103-2112-M-001-003-MY3.
V.B. and F.C. are supported by the Swiss National
Science Foundation (SNSF).
C.D.F and C.E.R. are funded through the NSF grant AST-1312329,
``Collaborative Research: Accurate cosmology with strong gravitational
lens time delays,'' and the {\it HST} grant GO-12889.
D.S. acknowledges funding support from a {\it {Back to Belgium}} grant from the Belgian Federal Science Policy (BELSPO).
T.T. thanks the Packard Foundation for generous support through a Packard
Research Fellowship, the NSF for funding through NSF grant AST-1450141,
``Collaborative Research: Accurate cosmology with strong gravitational lens time delays".
S.H. acknowledges support by the DFG cluster of excellence \lq{}Origin and Structure of the Universe\rq{} (\href{http://www.universe-cluster.de}{\texttt{www.universe-cluster.de}}).
L.V.E.K. is supported in part through an NWO-VICI career grant (project number 639.043.308).
P.J.M. acknowledges support from the U.S.\ Department of Energy under
contract number DE-AC02-76SF00515.
This paper is based on observations made with the NASA/ESA
{\it Hubble Space Telescope}, obtained at the Space Telescope Science
Institute, which is operated by the Association of Universities for
Research in Astronomy, Inc., under NASA contract NAS 5-26555. These
observations are associated with Programs \#12889 and \#9744.
Support for program \#12889 was provided by NASA through a grant from the Space Telescope Science Institute, which is operated by the Association of Universities for Research in Astronomy, Inc., under NASA contract NAS 5-26555.
Some of the data presented herein were obtained at the W.M. Keck
Observatory, which is operated as a scientific partnership among the
California Institute of Technology, the University of California and
the National Aeronautics and Space Administration. The Observatory was
made possible by the generous financial support of the W.M. Keck
Foundation.
The authors wish to recognize and acknowledge the very significant
cultural role and reverence that the summit of Mauna Kea has always
had within the indigenous Hawaiian community.  We are most fortunate
to have the opportunity to conduct observations from this mountain.

%-------------------------------------------------------------------------------
% bibliography:

\bibliography{he0435lensmodel}
\bibliographystyle{mnras}

%-------------------------------------------------------------------------------

\appendix

\section{Model Parameters}\label{app:params}
We show the marginalized parameter constraints for each of the SPEMD models in Table~\ref{tab:spemd_params} and for each of the composite models in Table~\ref{tab:comp_params}.

\renewcommand\tabcolsep{2pt}
\renewcommand*\arraystretch{1.5}
\begin{table*}
\caption{SPEMD Model Parameters \label{tab:spemd_params}}
\begin{minipage}{\linewidth}
\begin{tabular}{l|ccccccccc}
\hline
Parameter &
\multicolumn{9}{c}{Marginalized Constraints}
\\
\hline
 &
Fiducial &
Img+10 &
Arc+1,50src &
Arc+2,Im+10,50src &
AGNwht=0 &
AGNmask+1 &
AGNmask+2 &
5 pert. &
S\'{e}rsic
\\
\hline
$\theta_{\mathrm{E}}~(\arcsec)$\footnote{Spherical-equivalent Einstein radius} &
$1.182_{-0.002}^{+0.002}$ & $1.182_{-0.002}^{+0.002}$ &$1.169_{-0.001}^{+0.002}$ &$1.176_{-0.002}^{+0.002}$ &$1.180_{-0.002}^{+0.002}$ &$1.182_{-0.002}^{+0.002}$ &$1.181_{-0.002}^{+0.002}$ &$1.149_{-0.003}^{+0.003}$ &$1.181_{-0.002}^{+0.002}$
\\
$q$ &
$0.80_{-0.01}^{+0.01}$ &$0.81_{-0.01}^{+0.01}$ &$0.81_{-0.01}^{+0.01}$ &$0.81_{-0.01}^{+0.01}$ &$0.81_{-0.01}^{+0.01}$ &$0.80_{-0.01}^{+0.01}$ &$0.80_{-0.01}^{+0.01}$ &$0.81_{-0.01}^{+0.01}$ &$0.81_{-0.01}^{+0.01}$
\\
$\theta_{q}$ ($^{\circ}$) &
$-16.8_{-0.6}^{+0.5}$ &$-17.0_{-0.5}^{+0.5}$ &$-17.1_{-0.4}^{+0.5}$ &$-16.6_{-0.6}^{+0.5}$ &$-16.8_{-0.5}^{+0.5}$ &$-17.1_{-0.6}^{+0.6}$ &$-17.3_{-0.6}^{+0.5}$ &$-17.8_{-0.5}^{+0.5}$ &$-17.0_{-0.6}^{+0.6}$
\\
$\gamma^{\prime}$ &
$1.93_{-0.02}^{+0.02}$ &$1.95_{-0.03}^{+0.02}$ &$1.89_{-0.02}^{+0.02}$ &$1.91_{-0.01}^{+0.02}$ &$1.94_{-0.02}^{+0.02}$ &$1.94_{-0.02}^{+0.02}$ &$1.94_{-0.02}^{+0.02}$ &$1.93_{-0.01}^{+0.02}$ &$1.87_{-0.02}^{+0.03}$
\\
$\gext$ &
$0.030_{-0.003}^{+0.003}$ &$0.033_{-0.003}^{+0.003}$ &$0.032_{-0.002}^{+0.002}$ &$0.030_{-0.004}^{+0.003}$ &$0.033_{-0.003}^{+0.003}$ &$0.032_{-0.003}^{+0.003}$ &$0.031_{-0.003}^{+0.003}$ &$0.025_{-0.002}^{+0.002}$ &$0.026_{-0.003}^{+0.003}$
\\
$\theta_{\gamma}$ ($^{\circ}$) &
$63.7_{-2.2}^{+2.4}$ &$65.0_{-1.8}^{+1.9}$ &$57.7_{-1.6}^{+1.2}$ &$60.6_{-1.7}^{+2.1}$ &$63.6_{-1.9}^{+1.9}$ &$65.3_{-2.0}^{+1.9}$ &$65.4_{-2.0}^{+2.0}$ &$-88.5_{-1.3}^{+1.4}$ &$63.1_{-2.7}^{+2.7}$
\\
G1 $\theta_{\mathrm{E}}~(\arcsec)$ &
$0.37_{-0.03}^{+0.03}$ &$0.38_{-0.03}^{+0.02}$ &$0.48_{-0.02}^{+0.02}$ &$0.40_{-0.02}^{+0.02}$ &$0.39_{-0.03}^{+0.02}$ &$0.37_{-0.03}^{+0.03}$ &$0.37_{-0.03}^{+0.03}$ &$0.26_{-0.01}^{+0.01}$ &$0.35_{-0.03}^{+0.03}$
\\
\hline
\end{tabular}
\\
{\footnotesize Reported values are medians, with errors corresponding to the 16th and 84th percentiles.}
\\
{\footnotesize Angles are measured east of north.}
\end{minipage}
\end{table*}
\renewcommand*\arraystretch{1.0}
\renewcommand\tabcolsep{6pt}

\renewcommand*\arraystretch{1.5}
\begin{table*}
\caption{Composite Model Parameters \label{tab:comp_params}}
\begin{minipage}{\linewidth}
\begin{tabular}{l|ccc}
\hline
Parameter &
\multicolumn{3}{c}{Marginalized Constraints}
\\
\hline
 &
Composite &
Composite,AGNmask+1 &
Composite,Arcmask+1,50src
\\
\hline
Stellar M/L ($\mathrm{M_{\odot}/L_{\odot}}$)\footnote{M/L within $\theta_{\mathrm{E}}$ for rest-frame $V$ band.  The given uncertainties are only statistical and do not include systematic effects.  The stellar mass is calculated assuming $H_{0} = 70~\mathrm{km~s^{-1}~Mpc^{-1}}$, $\Omega_{m} = 0.3$, $\Omega_{\Lambda} = 0.7$, but changes in the cosmology affect the M/L by a negligible amount.} &
$2.5_{-0.1}^{+0.1}$ &$2.6_{-0.2}^{+0.2}$ &$2.3_{-0.1}^{+0.1}$
\\
NFW $\kappa_{0,h}$ &
$0.41_{-0.03}^{+0.03}$ &$0.36_{-0.03}^{+0.03}$ &$0.39_{-0.01}^{+0.01}$
\\
NFW $r_{s}~(\arcsec)$ &
$8.43_{-1.94}^{+0.58}$ &$9.43_{-0.94}^{+0.69}$ &$8.96_{-0.26}^{+0.28}$
\\
NFW $q$ &
$0.82_{-0.02}^{+0.01}$ &$0.81_{-0.02}^{+0.02}$ &$0.83_{-0.01}^{+0.01}$
\\
NFW $\theta_{q}$ ($^{\circ}$) &
$-18.4_{-0.7}^{+0.7}$ &$-18.6_{-0.7}^{+0.7}$ &$-19.7_{-0.6}^{+0.6}$
\\
$\gext$ &
$0.004_{-0.002}^{+0.003}$ &$0.003_{-0.002}^{+0.003}$ &$0.006_{-0.002}^{+0.002}$
\\
$\theta_{\gamma}$ ($^{\circ}$) &
$34.4_{-32.5}^{+22.9}$ &$44.6_{-36.2}^{+26.8}$ &$28.3_{-7.6}^{+6.0}$
\\
G1 $\theta_{\mathrm{E}}~(\arcsec)$ &
$0.33_{-0.03}^{+0.03}$ &$0.32_{-0.03}^{+0.03}$ &$0.42_{-0.02}^{+0.03}$
\\
\hline
\end{tabular}
\\
{\footnotesize Reported values are medians, with errors corresponding to the 16th and 84th percentiles.}
\\
{\footnotesize Angles are measured east of north.}
\end{minipage}
\end{table*}
\renewcommand*\arraystretch{1.0}

\section{Inverse Magnification Tensors}\label{app:invmagtens}
The components of the inverse magnification tensor are 
\begin{equation}\label{eq:invmagtens}
\mathcal{A}_{ij}(\bm{\theta}) = \frac{\partial {\beta_i}}{\partial {\theta_{j}}} 
\end{equation}
where $i=1,2$, $j=1,2$, $\bm{\beta}=(\beta_1,\beta_2)$ is the source plane coordinates, 
and $\bm{\theta}=(\theta_{1},\theta_{2})$ is the coordinates of the image plane (which is also the first lens plane, $\bm{\theta_1}=(\theta_{1_1}, \theta_{1_2})$).

The general multi-plane lens equation is
\begin{equation}\label{eq:mpfull}
\bm{\theta}_{j} = \bm{\theta}_{1} - \sum_{i=1}^{j-1} \beta_{ij} \bm{\alpha}_{i}(\bm{\theta}_{i}),
\end{equation}
where $\beta_{ij}$ is given by \eref{eq:beta} (note the difference between $\beta_{ij}$ with two indices and the source coordinates $\beta_i$ with one index).  This is the general form of \eref{eq:mp2a} and \eref{eq:mp2b}.  For $N$ lens planes, the source coordinates are $\bm{\beta}=\bm{\theta}_{N+1}$. For the case of two lens planes, as we have in our model, $\bm{\beta} = \bm{\theta}_{3}$.  We present the inverse magnification tensors at the positions of the lensed quasar images in Table~\ref{tab:invmag}.
While the inverse magnification tensor is symmetric for single-plane lensing, this is not true for multi-plane lensing. 

\renewcommand*\arraystretch{1.5}
\begin{landscape}
\begin{table}
\caption{Inverse Magnification Tensor \label{tab:invmag}}
\begin{minipage}{\linewidth}
\begin{tabular}{l|cccc}
\hline
Model &
A &
B &
C &
D
\\\hline
Fiducial &
$\arraycolsep2pt \begin{bmatrix} 0.732_{-0.016}^{+0.018} &-0.287_{-0.007}^{+0.006} \\-0.302_{-0.007}^{+0.007} &0.328_{-0.008}^{+0.008} \end{bmatrix}$ &
$\arraycolsep2pt \begin{bmatrix} -0.105_{-0.004}^{+0.004} &0.252_{-0.006}^{+0.007} \\0.281_{-0.007}^{+0.008} &0.857_{-0.021}^{+0.023} \end{bmatrix}$ &
$\arraycolsep2pt \begin{bmatrix} 0.832_{-0.017}^{+0.020} &-0.009_{-0.002}^{+0.002} \\-0.024_{-0.003}^{+0.003} &0.184_{-0.005}^{+0.005} \end{bmatrix}$ &
$\arraycolsep2pt \begin{bmatrix} -0.220_{-0.007}^{+0.008} &0.212_{-0.006}^{+0.007} \\0.249_{-0.007}^{+0.008} &0.850_{-0.022}^{+0.025} \end{bmatrix}$
\\
img+10pix &
$\arraycolsep2pt \begin{bmatrix} 0.743_{-0.020}^{+0.014} &-0.292_{-0.006}^{+0.007} \\-0.308_{-0.007}^{+0.009} &0.333_{-0.010}^{+0.008} \end{bmatrix}$ &
$\arraycolsep2pt \begin{bmatrix} -0.105_{-0.004}^{+0.005} &0.256_{-0.007}^{+0.006} \\0.286_{-0.009}^{+0.007} &0.876_{-0.026}^{+0.020} \end{bmatrix}$ &
$\arraycolsep2pt \begin{bmatrix} 0.846_{-0.022}^{+0.015} &-0.009_{-0.002}^{+0.002} \\-0.024_{-0.003}^{+0.003} &0.186_{-0.006}^{+0.004} \end{bmatrix}$ &
$\arraycolsep2pt \begin{bmatrix} -0.220_{-0.006}^{+0.009} &0.216_{-0.007}^{+0.006} \\0.254_{-0.009}^{+0.007} &0.870_{-0.027}^{+0.021} \end{bmatrix}$
\\
Arcmask+1pix,50x50src &
$\arraycolsep2pt \begin{bmatrix} 0.692_{-0.013}^{+0.013} &-0.274_{-0.005}^{+0.004} \\-0.292_{-0.006}^{+0.005} &0.313_{-0.006}^{+0.006} \end{bmatrix}$ &
$\arraycolsep2pt \begin{bmatrix} -0.099_{-0.003}^{+0.003} &0.236_{-0.005}^{+0.005} \\0.271_{-0.006}^{+0.007} &0.809_{-0.016}^{+0.018} \end{bmatrix}$ &
$\arraycolsep2pt \begin{bmatrix} 0.776_{-0.014}^{+0.014} &-0.010_{-0.002}^{+0.002} \\-0.029_{-0.002}^{+0.002} &0.178_{-0.004}^{+0.004} \end{bmatrix}$ &
$\arraycolsep2pt \begin{bmatrix} -0.207_{-0.006}^{+0.005} &0.185_{-0.004}^{+0.004} \\0.230_{-0.005}^{+0.006} &0.797_{-0.017}^{+0.019} \end{bmatrix}$
\\
Arcmask+2pix,img+10pix,50x50src &
$\arraycolsep2pt \begin{bmatrix} 0.711_{-0.011}^{+0.011} &-0.280_{-0.005}^{+0.005} \\-0.296_{-0.005}^{+0.005} &0.319_{-0.005}^{+0.005} \end{bmatrix}$ &
$\arraycolsep2pt \begin{bmatrix} -0.102_{-0.003}^{+0.003} &0.243_{-0.004}^{+0.005} \\0.274_{-0.005}^{+0.005} &0.830_{-0.014}^{+0.019} \end{bmatrix}$ &
$\arraycolsep2pt \begin{bmatrix} 0.805_{-0.013}^{+0.014} &-0.010_{-0.002}^{+0.002} \\-0.026_{-0.002}^{+0.003} &0.180_{-0.002}^{+0.003} \end{bmatrix}$ &
$\arraycolsep2pt \begin{bmatrix} -0.211_{-0.005}^{+0.004} &0.201_{-0.005}^{+0.006} \\0.241_{-0.005}^{+0.006} &0.820_{-0.015}^{+0.020} \end{bmatrix}$
\\
AGN mask weight=0 &
$\arraycolsep2pt \begin{bmatrix} 0.736_{-0.016}^{+0.017} &-0.291_{-0.007}^{+0.006} \\-0.307_{-0.007}^{+0.007} &0.332_{-0.008}^{+0.008} \end{bmatrix}$ &
$\arraycolsep2pt \begin{bmatrix} -0.103_{-0.004}^{+0.004} &0.255_{-0.007}^{+0.007} \\0.286_{-0.008}^{+0.008} &0.868_{-0.021}^{+0.022} \end{bmatrix}$ &
$\arraycolsep2pt \begin{bmatrix} 0.837_{-0.018}^{+0.019} &-0.007_{-0.002}^{+0.002} \\-0.023_{-0.003}^{+0.003} &0.185_{-0.005}^{+0.005} \end{bmatrix}$ &
$\arraycolsep2pt \begin{bmatrix} -0.219_{-0.008}^{+0.007} &0.211_{-0.007}^{+0.007} \\0.250_{-0.007}^{+0.008} &0.863_{-0.023}^{+0.024} \end{bmatrix}$
\\
AGN mask+1pix &
$\arraycolsep2pt \begin{bmatrix} 0.737_{-0.015}^{+0.016} &-0.290_{-0.006}^{+0.006} \\-0.306_{-0.007}^{+0.007} &0.334_{-0.007}^{+0.008} \end{bmatrix}$ &
$\arraycolsep2pt \begin{bmatrix} -0.106_{-0.004}^{+0.004} &0.256_{-0.006}^{+0.006} \\0.286_{-0.007}^{+0.008} &0.868_{-0.020}^{+0.021} \end{bmatrix}$ &
$\arraycolsep2pt \begin{bmatrix} 0.840_{-0.017}^{+0.017} &-0.008_{-0.002}^{+0.002} \\-0.023_{-0.003}^{+0.003} &0.187_{-0.004}^{+0.005} \end{bmatrix}$ &
$\arraycolsep2pt \begin{bmatrix} -0.223_{-0.007}^{+0.007} &0.216_{-0.006}^{+0.006} \\0.253_{-0.007}^{+0.007} &0.863_{-0.022}^{+0.023} \end{bmatrix}$
\\
AGN mask+2pix &
$\arraycolsep2pt \begin{bmatrix} 0.734_{-0.015}^{+0.016} &-0.290_{-0.006}^{+0.005} \\-0.306_{-0.007}^{+0.006} &0.334_{-0.007}^{+0.008} \end{bmatrix}$ &
$\arraycolsep2pt \begin{bmatrix} -0.105_{-0.004}^{+0.004} &0.256_{-0.006}^{+0.006} \\0.285_{-0.008}^{+0.007} &0.863_{-0.019}^{+0.022} \end{bmatrix}$ &
$\arraycolsep2pt \begin{bmatrix} 0.837_{-0.016}^{+0.018} &-0.008_{-0.002}^{+0.002} \\-0.023_{-0.003}^{+0.003} &0.186_{-0.005}^{+0.005} \end{bmatrix}$ &
$\arraycolsep2pt \begin{bmatrix} -0.223_{-0.007}^{+0.007} &0.216_{-0.006}^{+0.006} \\0.253_{-0.006}^{+0.008} &0.858_{-0.020}^{+0.023} \end{bmatrix}$
\\
5 perturbers &
$\arraycolsep2pt \begin{bmatrix} 0.707_{-0.011}^{+0.012} &-0.285_{-0.004}^{+0.004} \\-0.301_{-0.005}^{+0.005} &0.317_{-0.005}^{+0.006} \end{bmatrix}$ &
$\arraycolsep2pt \begin{bmatrix} -0.090_{-0.003}^{+0.003} &0.249_{-0.004}^{+0.006} \\0.277_{-0.005}^{+0.006} &0.826_{-0.012}^{+0.015} \end{bmatrix}$ &
$\arraycolsep2pt \begin{bmatrix} 0.802_{-0.012}^{+0.014} &-0.006_{-0.002}^{+0.002} \\-0.019_{-0.002}^{+0.002} &0.181_{-0.003}^{+0.004} \end{bmatrix}$ &
$\arraycolsep2pt \begin{bmatrix} -0.216_{-0.006}^{+0.005} &0.202_{-0.005}^{+0.005} \\0.237_{-0.005}^{+0.006} &0.841_{-0.013}^{+0.017} \end{bmatrix}$
\\
S\'{e}rsic profiles &
$\arraycolsep2pt \begin{bmatrix} 0.688_{-0.015}^{+0.019} &-0.271_{-0.007}^{+0.006} \\-0.285_{-0.008}^{+0.007} &0.309_{-0.008}^{+0.009} \end{bmatrix}$ &
$\arraycolsep2pt \begin{bmatrix} -0.098_{-0.004}^{+0.004} &0.236_{-0.006}^{+0.007} \\0.262_{-0.007}^{+0.008} &0.801_{-0.019}^{+0.025} \end{bmatrix}$ &
$\arraycolsep2pt \begin{bmatrix} 0.784_{-0.018}^{+0.022} &-0.008_{-0.002}^{+0.002} \\-0.022_{-0.003}^{+0.002} &0.173_{-0.004}^{+0.005} \end{bmatrix}$ &
$\arraycolsep2pt \begin{bmatrix} -0.205_{-0.008}^{+0.007} &0.198_{-0.007}^{+0.008} \\0.231_{-0.007}^{+0.008} &0.790_{-0.021}^{+0.027} \end{bmatrix}$
\\
Composite &
$\arraycolsep2pt \begin{bmatrix} 0.737_{-0.026}^{+0.034} &-0.281_{-0.010}^{+0.009} \\-0.295_{-0.011}^{+0.010} &0.315_{-0.010}^{+0.012} \end{bmatrix}$ &
$\arraycolsep2pt \begin{bmatrix} -0.104_{-0.007}^{+0.006} &0.251_{-0.008}^{+0.009} \\0.276_{-0.010}^{+0.011} &0.837_{-0.025}^{+0.028} \end{bmatrix}$ &
$\arraycolsep2pt \begin{bmatrix} 0.833_{-0.027}^{+0.028} &-0.016_{-0.015}^{+0.007} \\-0.029_{-0.016}^{+0.006} &0.180_{-0.006}^{+0.007} \end{bmatrix}$ &
$\arraycolsep2pt \begin{bmatrix} -0.209_{-0.010}^{+0.008} &0.212_{-0.008}^{+0.008} \\0.244_{-0.009}^{+0.010} &0.834_{-0.026}^{+0.029} \end{bmatrix}$
\\
Composite,AGN mask+1pix &
$\arraycolsep2pt \begin{bmatrix} 0.738_{-0.026}^{+0.025} &-0.291_{-0.010}^{+0.011} \\-0.304_{-0.011}^{+0.012} &0.329_{-0.012}^{+0.011} \end{bmatrix}$ &
$\arraycolsep2pt \begin{bmatrix} -0.103_{-0.006}^{+0.006} &0.257_{-0.010}^{+0.009} \\0.282_{-0.013}^{+0.011} &0.855_{-0.030}^{+0.028} \end{bmatrix}$ &
$\arraycolsep2pt \begin{bmatrix} 0.843_{-0.028}^{+0.027} &-0.008_{-0.002}^{+0.002} \\-0.022_{-0.003}^{+0.003} &0.183_{-0.007}^{+0.007} \end{bmatrix}$ &
$\arraycolsep2pt \begin{bmatrix} -0.217_{-0.010}^{+0.011} &0.221_{-0.008}^{+0.009} \\0.253_{-0.011}^{+0.011} &0.855_{-0.032}^{+0.030} \end{bmatrix}$
\\
Composite,Arcmask+1pix,50x50src &
$\arraycolsep2pt \begin{bmatrix} 0.705_{-0.013}^{+0.012} &-0.282_{-0.005}^{+0.005} \\-0.298_{-0.006}^{+0.006} &0.319_{-0.006}^{+0.006} \end{bmatrix}$ &
$\arraycolsep2pt \begin{bmatrix} -0.097_{-0.003}^{+0.003} &0.243_{-0.005}^{+0.004} \\0.275_{-0.006}^{+0.006} &0.814_{-0.014}^{+0.013} \end{bmatrix}$ &
$\arraycolsep2pt \begin{bmatrix} 0.795_{-0.014}^{+0.013} &-0.008_{-0.001}^{+0.001} \\-0.025_{-0.002}^{+0.002} &0.178_{-0.004}^{+0.003} \end{bmatrix}$ &
$\arraycolsep2pt \begin{bmatrix} -0.208_{-0.005}^{+0.005} &0.196_{-0.005}^{+0.004} \\0.236_{-0.006}^{+0.005} &0.811_{-0.015}^{+0.014} \end{bmatrix}$
\\
\hline
\end{tabular}
\\
\end{minipage}
\end{table}
\end{landscape}
\renewcommand*\arraystretch{1.0}

\bsp
\label{lastpage}
\end{document}